\documentclass[journal,twoside,web]{ieeecolor}
\usepackage{generic}
\usepackage{cite}
\usepackage{amsmath,amssymb,amsfonts}
\usepackage{algorithmic}
\usepackage{graphicx}
\usepackage{textcomp}
\usepackage{cite}
\usepackage{amsmath,amssymb,amsfonts}
\usepackage[english]{babel}
\usepackage{algorithmic}
\usepackage{graphicx}
\usepackage{textcomp}
\usepackage{optidef}
\usepackage{listings}
\usepackage[ruled,vlined]{algorithm2e}
\newtheorem{theorem}{Theorem}
\newtheorem{corollary}{Corollary}
\newtheorem{assumption}{Assumption}

\newtheorem{remark}{Remark}

\newtheorem{defn}{Definition}
\newtheorem{ex}{Example}

\usepackage{caption}
\usepackage{subcaption}

\usepackage{placeins}
\usepackage{mathtools}
\usepackage{lipsum}
\usepackage{braket}

\usepackage{bbding}
\usepackage{pifont}
\usepackage{wasysym}
\usepackage{amssymb}
\usepackage{multirow}

\newcommand{\revC}[1]{#1}
\newcommand{\revM}[1]{}

\usepackage{marginnote}

\newcommand{\Hs}{\mathcal{H}}

\newcommand{\Rspace}{\mathbb{R}}

\newcommand{\id}{\textbf{I}}

\newcommand{\Ss}{\mathcal{S}}

\newcommand{\X}{\mathcal{X}}

\newcommand{\Z}{\mathcal{Z}}
\newcommand{\U}{\mathcal{U}}
\newcommand{\R}{\mathcal{R}}
\newcommand{\Pp}{\mathcal{P}}

\newcommand{\C}{\mathcal{C}}

\newcommand{\DPhi}{\text{D}_{\Phi}}

\newcommand{\suc}{\text{Suc}}

\newcommand{\FSUSOL}{\Psi}
\newcommand{\FSUSCL}{\Phi}

\newcommand{\domFX}{\text{D}(\FSUSCL)}

\newcommand{\SIM}{\Theta}
\newcommand{\domSIM}{\text{D}_{\SIM}}

\usepackage{soul}

\def\BibTeX{{\rm B\kern-.05em{\sc i\kern-.025em b}\kern-.08em
    T\kern-.1667em\lower.7ex\hbox{E}\kern-.125emX}}
\begin{document}
\title{Reachability Analysis Using Hybrid Zonotopes and Functional Decomposition}
\author{Jacob A. Siefert, Trevor J. Bird, Andrew F. Thompson, Jonah J. Glunt, \\Justin P. Koeln, Neera Jain, and Herschel C. Pangborn
\thanks{Jacob A. Siefert, Andrew F. Thompson, Jonah J. Glunt, and Herschel C. Pangborn are with the Department of Mechanical Engineering, The Pennsylvania State University, University Park, PA 16802 USA (e-mails: jas7031@psu.edu; thompson@psu.edu; jglunt@psu.edu; \\hcpangborn@psu.edu).}
\thanks{Trevor J. Bird is with P. C. Krause and Associates, West Lafayette, IN 47906 USA (e-mail: tbird@pcka).}
\thanks{Justin P. Koeln is with the  Mechanical Engineering Department, University of Texas at Dallas, Richardson, TX 75080-3021 USA (e-mail: justin.koeln@utdallas.edu).}
\thanks{Neera Jain is with the School of Mechanical Engineering, Purdue University, West Lafayette, IN 47907 USA (e-mail: neerajain@purdue.edu).}
}

\maketitle
\thispagestyle{empty} 

{
\begin{abstract}

This paper proposes methods for reachability analysis of nonlinear systems in both open loop and closed loop with advanced controllers. The methods combine hybrid zonotopes, a construct called a state-update set, functional decomposition, and special ordered set approximations to enable linear growth in reachable set memory complexity with time and linear scaling in computational complexity with the system dimension. Facilitating this combination are new identities for constructing nonconvex sets that contain nonlinear functions and for efficiently converting a collection of polytopes from vertex representation to hybrid zonotope representation. Benchmark numerical examples from the literature demonstrate the proposed methods and provide comparison to state-of-the-art techniques.

\end{abstract}}
\section{Introduction}
\label{sect:introduction}

 {
 Reachability analysis---the process of calculating reachable sets---is used to evaluate system performance and ensure constraint satisfaction in safety-critical applications. However, the scalability of existing approaches for nonlinear systems is limited by their nonconvexity and the computational complexity that this induces. \revM{R3.1, R3.2}\revC{Several methods have been proposed to calculate reachable sets for continuous-time nonlinear systems, including Hamilton-Jacobi reachability \cite{Bansal2017HamiltonJacobiRA,Bui2021RealTimeHR,HJreachDecomp2018,FastHJreachApprox2016,OPTNL_mitchell2005time} and monotonicity-based techniques \cite{MONO_angeli2003monotone,MONO_ramdani2008reachability,CooganMixedMonotone}.} This paper addresses \emph{set propagation techniques}, which calculate reachable sets by recursively propagating successor sets \cite{Althoff_SetPropagationTechniques_2021}. 

\subsubsection{Gaps in literature} Set propagation techniques for continuous-time systems often resemble their discrete-time counterparts, as the former often still propagate reachable sets over discrete time intervals \cite{Althoff_ReachSafetyAutonomousCars_2010}. For this reason, there are many shared challenges in calculating reachable sets for continuous-time and discrete-time systems. Abstraction of the state space helps to mitigate the computational challenges of generating continuous-time and discrete-time reachable sets by approximating nonlinear functions with affine functions over partitioned regions of the domain. \revM{R8.6}\revC{In some cases, higher-order polynomial abstractions compatible with specific set representations are possible \cite{ConPolyZono_Althoff,SparsePolyZono_Althoff}. A collection of polyhedral over-approximations of the nonlinear dynamics can be generated by combining bounds for the error associated with higher-order terms with affine abstractions.} State-space abstractions can either be generated in a time-invariant \cite{ABS_INV_asarin2007hybridization,ABS_INV_asarin2003reachability} or time-varying manner \cite{ABS_TV_althoff2008reachability}. The primary challenge with time-invariant abstractions is that they exhibit hybrid behaviour, which can cause exponential growth in the number of convex reachable sets as a result of taking intersections across guards associated with the domain of each abstraction \cite{alur_algorithmic_1995,bemporad2003modeling,Althoff_SetPropagationTechniques_2021}. 

Recent work by the authors has proposed scalable approaches for reachability analysis of discrete-time hybrid systems using a new set representation called the hybrid zonotope, a construct called the state-update set, and efficient identities for using the state-update set to calculate successor and precursor sets \cite{SiefertHybSUS,Bird_HybZono,Bird_HybZonoMPC,BIRDthesis_2022,Bird_HybZonoUnionComp}. 
These methods were extended for over-approximated reachability analysis of nonlinear systems by representing special ordered sets as hybrid zonotopes \cite{Siefert2022}, where open-loop state-update sets were combined with a set representation of the controller called the state-input map to construct closed-loop state-update sets. Figure~\ref{fig:MethodOutline_CL} depicts this process of combining an open-loop state-update set $\Psi$ with a state-input map $\Theta$ to obtain a closed-loop state-update set $\Phi$, which can then be applied iteratively to calculate reachable sets. While construction of an open-loop state-update set and state-input map were detailed for a particular numerical example in \cite{Siefert2022}, a generalized method was not provided.

\begin{figure*}
    \centering
    \includegraphics[width=.7\textwidth]{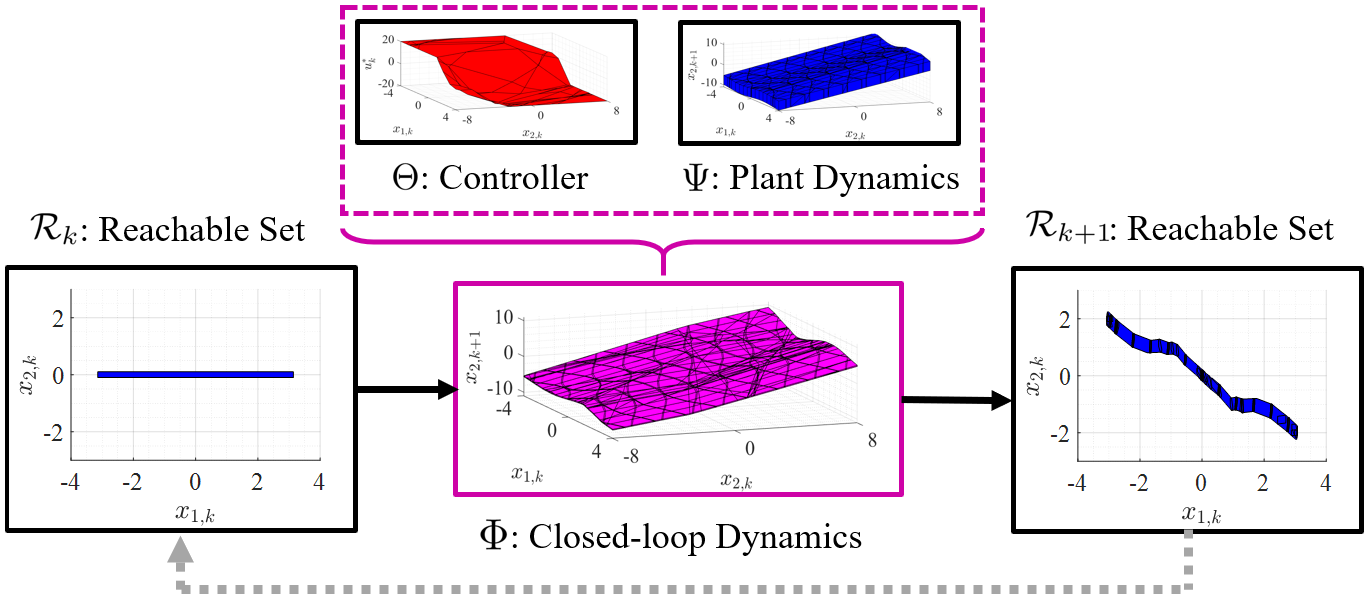}
    \caption{The closed-loop successor set identity uses a set-based representation of the closed-loop dynamics, called the closed-loop state-update set $\Phi$, to generate the one-step forward reachable set $\R_{k+1}$ from $\R_{k}$. The closed-loop state-update set is created by combining sets representing the open-loop dynamics and a state-feedback controller, called the open-loop state-update set $\Psi$ and the state-input map $\Theta$, respectively.}
    \label{fig:MethodOutline_CL}
\end{figure*}

\subsubsection{Contribution} 
\revM{R6.1, R6.2, R8.3}\revC{This paper provides efficient identities to construct sets that contain general nonlinear functions. The identities are compatible with hybrid zonotopes and can be used to generate both open-loop state-update sets and state-input maps, which can then be used within successor set identities to calculate over-approximations of reachable sets.} The method decomposes functions with an arbitrary number of arguments into functions with only one or two scalar arguments, thus avoiding the so-called curse of dimensionality (i.e., exponential growth in memory complexity with respect to the state dimension). \revM{R3.5,  R8.3}\revC{Complementary results for hybrid zonotopes are provided, including efficient conversion from collections of vertex representation polytopes and efficient over-approximation of common binary functions.} Theoretical results are applied to several numerical examples. The first performs reachability of an inverted pendulum with a neural network controller. This aligns with a benchmark problem from \cite{ARCH22:AINNCS,ARCH23:AINNCS}, facilitating comparison to other state-of-the-art methods. While hybrid zonotopes have previously been used for reachability analysis of neural networks \cite{Zhang_VerNNusingHZ_2022,JKJR_NNusingHZ_2022}, the methods in this paper achieve improved scalability and go beyond the scope of previous work by analyzing the coupling of a neural network controller to a nonlinear plant. The second numerical example performs reachability of a high-dimensional logical system from \cite{alanwar2023polynomial} to demonstrate applications to logical systems and scalability with the number of states. The final numerical example performs reachability of the Vertical Collision Avoidance System (VCAS) from \cite{ARCH22:AINNCS,ARCH23:AINNCS} to demonstrate how the proposed techniques can address a complex combination of neural networks and logical operations, while other state-of-the-art methods require simplifying assumptions.

\subsubsection{Outline} The remainder of this paper is organized as follows. Section \ref{sec:Prelims} provides notation and basic definitions. Section \ref{sect:SUS_reach} defines open-loop and closed-loop state-update sets and presents an identity to construct a closed-loop state-update set from an open-loop state-update set and a state-input map. This section also provides a generalized method for efficiently constructing open-loop state-update sets and state-input maps by leveraging functional decomposition. Section \ref{sec:HybZonoSUS} shows how to generate over-approximated sets of nonlinear functions by efficiently converting a collection of vertex-representation polytopes into a single hybrid zonotope. Section \ref{sect:Examples} applies the theoretical results to several numerical examples. 

\section{Preliminaries and Previous Work}
\label{sec:Prelims}

\subsection{Notation}
Matrices are denoted by uppercase letters, e.g., $G\in\Rspace^{n\times n_g}$, and sets by uppercase calligraphic letters, e.g., $\mathcal{Z}\subset\Rspace^{n}$. Vectors and scalars are denoted by lowercase letters. The $i^{th}$ column of a matrix $G$ is denoted by $G_{(\cdot,i)}$. Commas in subscripts are used to distinguish between properties that are defined for multiple sets, e.g., $n_{g,z}$ describes the complexity of the representation of $\mathcal{Z}$ while $n_{g,w}$ describes the complexity of the representation of $\mathcal{W}$. The topological boundary of a set $\mathcal{Z}$ is denoted by $\partial\mathcal{Z}$ and its interior by $\mathcal{Z}^{\circ}$. The $n$-dimensional unit hypercube is denoted by $\mathcal{B}_{\infty}^n=\left\{x\in\Rspace^{n}~|~\|x\|_{\infty}\leq1\right\}$. The set of all $n$-dimensional binary vectors is denoted by $\{-1,1\}^{n}$ and the interval set between a lower bound $b_{l}$ and an upper bound $b_{u}$ is denoted by $[b_{l},b_{u}]$. Matrices of all $0$ and $1$ elements are denoted by $\mathbf{0}$ and $\mathbf{1}$, respectively, of appropriate dimension and \revC{$\id$ denotes the identity matrix}\revM{R3.6}. The $i^{th}$ row of the identity matrix is denoted by $e_i$. The concatenation of two column vectors to a single column vector is denoted by $(g_1,\:g_2)=[g_1^T\:g_2^T]^T$ and the rounding of a scalar $s$ to the next largest integer is denoted by $\lceil s \rceil$. 

Given the sets $\mathcal{Z},\mathcal{W}\subset\Rspace^{n},\:\mathcal{Y}\subset\Rspace^{m}$, and matrix $R\in\Rspace^{m\times n}$, the linear transformation of $\mathcal{Z}$ by $R$ is $R\mathcal{Z}=\{Rz~|~z\in\mathcal{Z}\}$, the Minkowski sum of $\mathcal{Z}$ and $\mathcal{W}$ is $\mathcal{Z}\oplus\mathcal{W}=\{z+w~|~z\in\mathcal{Z},\:w\in\mathcal{W}\}$, the generalized intersection of $\mathcal{Z}$ and $\mathcal{Y}$ under $R$ is $\mathcal{Z}\cap_R\mathcal{Y}=\{z\in\mathcal{Z}~|~Rz\in\mathcal{Y}\}$, the Cartesian product of $\mathcal{Z}$ and $\mathcal{Y}$ is $\mathcal{Z}\times\mathcal{Y}=\{(z,y)|~z\in\mathcal{Z},\:y\in\mathcal{Y}\}$. 

\subsection{Set representations}
\label{sec:HybZono}

\begin{defn}
    \label{def-VrepPoly}
    A set $\Pp\in\Rspace^n$ is a convex polytope if it is bounded and $\exists\ V\in\Rspace^{n \times n_v}$ with $n_v<\infty$ such that
    \begin{align}
    \nonumber
        \Pp = \left\{ V\lambda\ |\ \lambda_j\geq 0\ \forall j\in \{1,...,n_v\},\ \textbf{1}^T\lambda=1\ \right\}\:.
    \end{align}
\end{defn}

A convex polytope is the convex hull of a finite set of vertices given by the columns of $V$. A convex polytope defined using a matrix of vertices is said to be given in \emph{vertex-representation} (V-rep). \revM{R8.1}\revC{The memory complexity and computational complexity of many key set operations scales poorly  when V-rep is used \cite[Table~1]{Althoff_SetPropagationTechniques_2021}, often limiting analysis to systems with few states. The methods proposed in  this paper use functional decomposition to represent high-dimensional systems using unary or binary functions, thus we only use vertex representation in one- or two-dimensional space. The resulting V-rep polytopes are then converted to hybrid zonotopes, which are combined in a higher-dimensional space.}

\begin{defn}\label{def-hybridZono} \cite[Def. 3]{Bird_HybZono}
The set $\mathcal{Z}_h\subset\Rspace^n$ is a \emph{hybrid zonotope} if there exist $G^c\in\Rspace^{n\times n_{g}}$, $G^b\in\Rspace^{n\times n_{b}}$, $c\in\Rspace^{n}$, $A^c\in\Rspace^{n_{c}\times n_{g}}$, $A^b\in\Rspace^{n_{c}\times n_{b}}$, and $b\in\Rspace^{n_c}$ such that {\small
    \begin{equation}\label{def-eqn-hybridZono}
        \mathcal{Z}_h = \left\{ \left[G^c \: G^b\right]\left[\begin{smallmatrix}\xi^c \\ \xi^b \end{smallmatrix}\right]  + c\: \middle| \begin{matrix} \left[\begin{smallmatrix}\xi^c \\ \xi^b \end{smallmatrix}\right]\in \mathcal{B}_\infty^{n_{g}} \times \{-1,1\}^{n_{b}}, \\ \left[A^c \: A^b\right]\left[\begin{smallmatrix}\xi^c \\ \xi^b \end{smallmatrix}\right] = b \end{matrix} \right\}\:.
\end{equation}}
\vskip \baselineskip
\end{defn}

 A hybrid zonotope is the union of $2^{n_b}$ constrained zonotopes corresponding to all combinations of binary factors. The hybrid zonotope is given in \textit{hybrid constrained generator representation} and the shorthand notation of $\mathcal{Z}_h=\langle G^c,G^b,c,A^c,A^b,b\rangle\subset\Rspace^n$ is used to denote the set given by \eqref{def-eqn-hybridZono}. Continuous and binary \emph{generators} refer to the columns of $G^c$ and $G^b$, respectively. A hybrid zonotope with no binary factors is a constrained zonotope, $\mathcal{Z}_c=\langle G,c,A,b\rangle\subset\Rspace^n$, and a hybrid zonotope with no binary factors and no constraints is a zonotope, $\mathcal{Z}=\langle G,c\rangle\subset\Rspace^n$. Identities and time complexities of linear transformations, Minkowski sums, generalized intersections, and generalized half-space intersections are reported in \cite[Section 3.2]{Bird_HybZono}. An identity and time complexity for Cartesian products is given in \cite{BIRDthesis_2022}. Methods for removing redundant generators and constraints of a hybrid zonotope were explored in \cite{Bird_HybZono} and developed further in \cite{BIRDthesis_2022}.

\subsection{Graphs of functions}

 A set-valued mapping $\phi: \DPhi \rightarrow\mathcal{Q}$ assigns each element $p\in\text{D}_{\Phi}$ to a subset, potentially a single point, of $\mathcal{Q}$. We refer to the set $\Phi = \{(p,q)\ |\ p\in\DPhi,\ q\in\phi(p)\subseteq\mathcal{Q} \}$ as a \emph{graph of the function} $\phi$. The graph of a function consists of ordered pairs of a function over a given domain. The set $\text{D}_{\Phi}$ is referred to as the \emph{domain set} of $\Phi$ and can be chosen by a user as the set of inputs of interest. \revM{R5.0}\revC{Graphs of functions, and the related concept of \emph{relations}, have been used for reachability analysis of nonlinear and hybrid systems, e.g., \cite{Morari_GOF,Sankaranarayanan_1,Sankaranarayanan_2}.} 
 
The authors have provided identities that leverage graphs of functions for reachability of hybrid system \cite{SiefertHybSUS}, set-valued state estimation \cite{Siefert2023_CDC}, and reachability of nonlinear system \cite{Siefert2022}. Some of the latter are included in Section \ref{sect:SUS_reach} (\textbf{Theorem~\ref{th:OneStep_F_OL}} and \textbf{Theorem~\ref{th:OneStep_F_CL}}) for completeness and to motivate the proposed identities for efficient construction of graphs of functions for nonlinear systems. The proposed methods for constructing graphs of nonlinear functions leverage hybrid zonotopes, special ordered set approximations and functional decomposition.

\subsection{Successor sets}
\label{sec:SuccessorIntro}

Consider a class of discrete-time nonlinear dynamics given by $f:\Rspace^{n} \times \Rspace^{n_u}\rightarrow\Rspace^n$
\begin{align} \label{eqn-genNLdyn}
    x_{k+1} = f(x_{k},u_{k}) \:,
\end{align}%
with state and input constraint sets given by $\mathcal{X}\subset\Rspace^n$ and $\mathcal{U}\subset\Rspace^{n_u}$, respectively. The $i^{th}$ row of $f(x_{k},u_{k})$ is a scalar-valued function and denoted by $f_i(x_{k},u_{k})$. Disturbances are omitted for simplicity of exposition, although the results in this paper extend to systems with disturbances. Because hybrid zonotopes are the set representation of interest for this paper and are inherently bounded, Assumption~\ref{ass:bounded} is made.

\begin{assumption}
\label{ass:bounded}
For all $(x,u)\in\X\times\U$, $||f(x,u)||<\infty$.
\end{assumption}

The successor set is defined as follows.

\begin{defn}
The \emph{successor set} from $\mathcal{R}_k\subseteq\mathcal{X}$ with inputs bounded by $\U_k\subseteq\U$ is given by 
\begin{align}\label{eqn-Suc}
    \suc(\R_{k},\U_k)\equiv\left\{\begin{matrix}
        f(x,u)
        \mid\:
        x\in\mathcal{R}_{k},\: u\in\U_k
    \end{matrix}
    \right\}\:.
\end{align}%
\vskip \baselineskip
\end{defn}
Forward reachable sets from an initial set can be found by recursion of successor sets \eqref{eqn-Suc}, i.e., $ \R_{k+1} =\suc(\R_{k},\U_k)$.

\subsection{Special ordered sets}
\label{sec:SOSintro}

Special Ordered Set (SOS) approximations, a type of piecewise-affine (PWA) approximations, were originally developed to approximate solutions of nonlinear optimization programs \cite{beale1970_SOS}. We define an SOS approximation equivalently to \cite[Section~1.2]{leyffer2008branch}. An incidence matrix is introduced to mirror the structure given to collections of V-rep polytopes in \textbf{Theorem~\ref{thm-SOS2HYBZONO}}.

\begin{defn}[SOS Approximation]
\revM{R1.4, R3.7, R5.4} \revC{An SOS approximation $\Ss$ of a scalar-valued function $g(x): \Rspace^n \rightarrow \Rspace$ is the union of $N$ polytopes, i.e., $\mathcal{S} = \cup_{i=1}^{N} \Pp_i$. The collection of polytopes is defined by a vertex matrix $V=[v_1 \, \cdots \, v_{n_v}]\in\Rspace^{(n+1)\times n_v}$, where $v_i = (x_i,g(x_i))\:,\ \forall i$, and a corresponding incidence matrix $M\in\Rspace^{n_v \times N}$ with entries $M_{(j,i)}\in\{0,1\}\:,\ \forall \:i,j$, such that 
{\small
    \begin{align}
    \label{defn-eqn-SOS-Pi}
        &\Pp_i=\left\{ V\lambda \bigg| \begin{array}{cc}
             \lambda_j \in \begin{cases} \begin{bmatrix} 0\:, & 1\end{bmatrix}, & \text{if}\ j\in \{ k\ |\ M_{(k,i)}=1 \}\: \\ \{0\}, & \text{if}\ j\in \{ k\ |\ M_{(k,i)}=0 \} \end{cases},  \\
             \mathbf{1}^T_{n_v} \lambda = 1 
        \end{array} \right\}\:,\\
        \label{defn-eqn-SOS-P-notouchy}
        &([\mathbf{\id}_{n}\ \mathbf{0}]\Pp_i)^{\circ}\cap [\mathbf{\id}_{n}\ \mathbf{0}]\Pp_j = \emptyset\:, \forall i \neq j \:, \text{ and} \\
        \label{defn-eqn-SOS-simplex}
        &\mathbf{1}^T M_{(\cdot,i)} \leq n+1\:,\ \forall\ i\in\{1,...,N\}\:.
    \end{align}}}%
\end{defn}

\revC{The set of points $x_i,\: i\in\{1,...,n_v\}$ are referred to as the \textit{sampling} of the first $n$ dimensions. Each polytope $\Pp_i$ given by \eqref{defn-eqn-SOS-Pi} is the convex hull of the vertices given by $V_{(\cdot,i)}$ corresponding to the index of all entries of $M_{(\cdot,i)}$ that equal $1$. The constraint \eqref{defn-eqn-SOS-P-notouchy} enforces that none of the simplices' interiors ``overlap'' while allowing for sharing of topological boundaries. The constraint \eqref{defn-eqn-SOS-simplex} enforces that $\Pp_i$ will be at most an $n$-dimensional simplex ($\mathbf{1}^TM_{(\cdot,i)} = n+1$) and a lower dimensional simplex otherwise ($\mathbf{1}^TM_{(\cdot,i)} < n+1$).}

\begin{ex}
\label{ex-sinSOS}
\revM{R1.3}Consider $y=\sin(x)$ for $x\in[-4,4]$. An SOS approximation with 21 evenly spaced breakpoints is given by vertex matrix $V$ and incidence matrix $M$ as

\begin{align}
\nonumber
    V &= \begin{bmatrix}
    -4 & -3.6 & -3.2 & \dots & 4\\ 
    \sin(-4) & \sin(-3.6) & \sin(-3.2) & \dots & \sin(4) 
    \end{bmatrix} \:,\ \text{and}\\
\nonumber
    M &= \begin{bmatrix}
        \id_{20}\\
        \mathbf{0}_{1 \times 20}
    \end{bmatrix} + \begin{bmatrix}
        \mathbf{0}_{1 \times 20}\\
        \id_{20}
    \end{bmatrix}\:.
\end{align}%
The first column of the incidence matrix
\begin{align}
\nonumber
    M_{(1,:)} = \begin{bmatrix}
        1 & 1 & 0 & \cdots & 0
    \end{bmatrix}^T
\end{align}%
corresponds to one section of the SOS approximation for the domain $x\in[-4,\ -3.6]$, which is a 1-dimensional simplex.
\end{ex}

\subsection{Functional decomposition}
\label{sec:funcDecomp}
 Functions can be decomposed into unary and binary functions with one or two scalar arguments, respectively. Constructing SOS approximations of the decomposed functions avoids exponential growth with respect to the argument dimension \cite{leyffer2008branch}. A function $h(x):\Rspace^n\rightarrow\Rspace^m$ is decomposed by introducing intermediate variables
\begin{align}
    \label{eqn-decomp}
    w_j & =
    \begin{cases}
    x_j \:, & \text{if } j=1,...,n\:,\\
    h_j(w_{j1}\{,w_{j2}\}) \:, & \text{if } j=n+1,...,n+K\:,
    \end{cases}
\end{align}
where $j1,j2<j$, giving
\begin{align}
    \nonumber
    h(x) &= \begin{bmatrix}
    w_{n+K-m+1}\\
    \vdots\\
    w_{n+K}
    \end{bmatrix} \:.
\end{align}%
The first $n$ assignments directly correspond to the $n$ elements of the argument vector $x$, assignments $n+1,...,n+K$ are defined by the unary function or binary function $h_j$, and the final $m$ assignments are associated with $h(x)$. In the case that $h_j$ is unary, the second argument is omitted.

\begin{remark}[Affine Decompositions]
\label{remark-FD-MoreThanTwoAllowed4Affine}
    Because hybrid zonotopes are closed under linear transformation, functional compositions are also allowed to admit functions $h_j(\cdot)$ that have more than two arguments, provided that $h_j(\cdot)$ is an affine function of lower-indexed variables.
\end{remark}

\begin{ex}
    \revM{R3.9}\revC{Consider the inverted pendulum dynamics given by
\begin{align}
    \label{eqn-PendulumContinuous}
    \begin{bmatrix}
    \Dot{x}_{1}\\
    \Dot{x}_{2}
    \end{bmatrix} = 
    \begin{bmatrix}
    x_2\\
    \frac{g}{l}\sin(x_1) + \frac{u}{I}\\
    \end{bmatrix}\:,
\end{align}
with gravity $g=10$, length $l=1$, mass $m=1$, and moment of inertia $I=ml^2=1$. The continuous-time nonlinear dynamics are discretized with time step $h=0.1$ using a $2^{nd}$-order Taylor polynomial $\mathcal{T}_2(x_k)$ given by
{\small
\begin{align} 
    \label{eqn-T2}
    \mathcal{T}_2(x_k) &= \left\{
    \begin{bmatrix}
    x_{1,k} +  \frac{x_{2,k}}{10}+\frac{\sin(x_{1,k})}{20}+\frac{u_k}{200}\\
    x_{2,k} + \sin(x_{1,k}) + \frac{x_{2,k}\cos(x_{1,k})}{20} + \frac{u_k}{10}
    \end{bmatrix}\right\}\:.
\end{align}}%
A functional decomposition of $\mathcal{T}_2(x_k)$ is shown in Table~\ref{tab:PendulumDecomp}. For a chosen domain $(x_{1,k},x_{2,k}.u_{k})\in\text{D}_\Hs=\text{D}_1 \times \text{D}_2 \times \text{D}_3$, bounds on the intermediate and output variables can be found by domain propagation via interval arithmetic.
\renewcommand{\arraystretch}{1.25} 
\begin{table}[]
\centering
\caption{Functional decomposition of $\mathcal{T}_2$ \eqref{eqn-T2} with $K=5$ compositions.}
\begin{tabular}{l|c|c}
$w_j(w_{j1},...)$           & $h_\ell$                                                  & $\text{D}_\ell$ \\ \hline \hline
$w_1 = x_{1,k}$ &                                                          & $[-4, 4]$    \\
$w_2 = x_{2,k}$ &                                                          & $[-8, 8]$    \\
$w_3 = u_k$     &                                                          & $[-20, 20]$  \\
$w_4(w_1)$           & $\sin(w_1)$                                              & $[-1, 1]$    \\
$w_5(w_1)$           & $\cos(w_1)$                                              & $[-1, 1]$    \\ 
$w_6(w_5,w_2)$           & $w_5 w_2$                                                & $[-8, 8]$    \\
{$w_7(w_1,w_2,w_3,w_4)$}           & {$w_1 + \frac{w_2}{10} + \frac{w_3}{200} +\frac{w_4}{20}$} &
$[-4.95, 4.95]$  \\
{$w_8(w_2,w_4,w_3,w_6)$}           & {$w_2+w_4+\frac{w_3}{10}+\frac{w_6}{20}$}                  & $[-11.4, 11.4]$
\end{tabular}
\label{tab:PendulumDecomp}
\end{table}
\renewcommand{\arraystretch}{1}} 
\end{ex}

\begin{remark}[Existence of a Functional Decomposition] \revM{R3.5, R3.14, R5.5}\revC{The Kolmogorov Superposition Theorem \cite{KolmogorovTheoremRelevant} proves that a continuous function $f(\cdot)$ defined on the $n$-dimensional hypercube can be represented as the sum and superposition of continuous functions of only one variable. This result allows us to analyze the scalability of the proposed approach in Section~\ref{sec:AvoidCoD}, although functional decompositions are system-specific, not unique, and not always obvious to find. Fortunately, decompositions of the form in \eqref{eqn-decomp} are readily available for large classes of functions, such as those containing basic operators (e.g., addition, subtraction, multiplication, division), polynomials, trigonometric functions, and boolean functions (e.g., AND). Additionally, they can often be obtained by analyzing the order of operations within an expression, although a decomposition strictly based on the order of operations may not be most concise or useful. 
} 
\end{remark}
\section{Reachability via State-update Sets}
\label{sect:SUS_reach}

This section first introduces the open-loop state-update set (Definition \ref{def:FSUS_OL}), which encodes all possible state transitions of \eqref{eqn-genNLdyn} over a user-specified domain of states and inputs, and is used to calculate successor sets over discrete time steps (\textbf{Theorem \ref{th:OneStep_F_OL}}). Then, after defining a state-input map as all possible inputs of a given control law over a user-specified domain of states (Definition \ref{def:StateInputMap}), the set of possible state transitions of the closed-loop system is constructed by combining the state-input map and the open-loop state-update set (\textbf{Theorem \ref{thm-CLSUSfromOL}}). It is shown how this closed-loop state-update set can be used to calculate successor sets of the closed-loop system (\textbf{Theorem \ref{th:OneStep_F_CL}}). Then, a general method to construct complex nonlinear sets using functional decomposition, later used to construct open-loop state update sets and state-input maps, is provided (\textbf{Theorem \ref{thm-FunctionCompSets}}). Finally, the effect of over-approximations on the theoretical results of this section is addressed (\textbf{Corollary \ref{co-SusOA_ReachOA}}).

\begin{defn}
\label{def:FSUS_OL}
The \textit{open-loop state-update set} $\FSUSOL\subseteq\Rspace^{2n+n_u}$ is defined as 
\begin{align}
    \label{eq:ForwardSUS}
    \FSUSOL \equiv \left\{\ \begin{bmatrix}
    x_k\\
    u\\
    x_{k+1}
    \end{bmatrix}\ \bigg |\ \begin{array}{c}
         x_{k+1} \in \suc (\{x_{k}\},\{u\}),\\
         (x_k,u) \in \text{D}_{\Psi}
    \end{array} \right\}\:.
\end{align}
\vskip \baselineskip
\end{defn}%
We refer to $\text{D}_{\Psi}\subset\Rspace^{n+n_u}$ as the \textit{domain set} of $\FSUSOL$, typically chosen as the region of interest for analysis.

\begin{theorem}\label{th:OneStep_F_OL}\cite[Theorem 1]{Siefert2022}
Given sets of states $\R_k\subseteq\Rspace^n$ and inputs $\U_k\subseteq\Rspace^{n_u}$, and an open-loop state-update set $\FSUSOL$, if $\R_{k}\times\U_k\subseteq \text{D}_{\Psi}$, then the \textit{open-loop successor set} is given by
\begin{align}
    \label{eq:1stepForward_OL}
    \suc(\R_k,\U_k) &= \begin{bmatrix}
        \textbf{0} & \id_{n}
        \end{bmatrix}\big(\FSUSOL\cap_{[\id_{n+n_u}~\mathbf{0}]} (\R_{k}\times\U_k)\big) \:.
\end{align}
\end{theorem}

The containment condition in \textbf{Theorem \ref{th:OneStep_F_OL}}, $\R_{k}\times\U_k\subseteq \text{D}_{\Psi}$, is not restrictive as modeled dynamics are often only valid over some region of states and inputs, which the user may specify as $\text{D}_{\Psi}=\X\times\U$.

Consider a set-valued function $\C(x_k)$ corresponding to a state-feedback controller, such that $\C(x_k)$ is the set of all possible inputs that the controller may provide given the current state, $x_k$. For example, for a linear feedback control law given by $u(x_k)=Kx_k$ with no actuator uncertainty, $\C(x_k) = \{Kx_k\}$ would be a single vector. In the case of a linear feedback control law with actuator uncertainty given by $u=Kx_k+\delta_u$ where $\delta_u \in \Delta_u$, we would have $\C(x_k) = \{Kx_k+\delta_u\ |\ \delta_u\in\Delta_u\}$. The \textit{state-input map} encodes the feedback control law given by $\C(x_k)$ as a set over a domain of states.
\begin{defn} \label{def:StateInputMap}
The \textit{state-input map }is defined as $\SIM \equiv \{(x_k,u)~|~u\in\C(x_k),\ x_k\in\domSIM \}$, where $\domSIM$ is the \emph{domain set} of $\SIM$.
\end{defn}

Next, the closed-loop state-update set under a controller given by $\C(x_k)$ is defined. Then it will be shown how to construct a closed-loop state-update set given an open-loop state-update set and a state-input map.

\begin{defn}
\label{def:FSUS_CL}
 The \textit{closed-loop state-update set} $\FSUSCL\subseteq\Rspace^{2n}$ for a controller given by $\C(x_k)$ is defined as 
\begin{align}
    \label{eq:ForwardSUS_CL}
    \FSUSCL \equiv \left\{ \begin{bmatrix}
    x_k\\
    x_{k+1}
    \end{bmatrix}\ \bigg |\ \begin{array}{c}
         x_{k+1} \in \suc \left(\{x_{k}\},\C(x_k)\right),\\
         x_k \in \text{D}_{\Phi}
    \end{array} \right\}\:,
\end{align}
\vskip \baselineskip
\end{defn}%
where $\text{D}_{\Phi}\subset\Rspace^{n}$ is the \textit{domain set} of $\FSUSCL$.
\begin{theorem} \label{thm-CLSUSfromOL} \cite[Theorem 2]{Siefert2022} Given an open-loop state-update set $\FSUSOL$ and state-input map $\SIM$, the closed-loop state-update set $\FSUSCL$ with\revM{R1.1} \revC{$\text{D}_{\Phi}=\begin{bmatrix}
\id_n~\mathbf{0}
\end{bmatrix}\left(\text{D}_{\Psi}\cap\SIM\right)$} is given by
\begin{align}
    \label{eqn-CLfromOL}
    \FSUSCL = \begin{bmatrix}
    \id_n & \mathbf{0} & \mathbf{0}\\
    \mathbf{0} & \mathbf{0} & \id_n\\
    \end{bmatrix} \left( \FSUSOL \cap_{\begin{bmatrix}
    \id_{n+n_u}~\mathbf{0}
    \end{bmatrix}} \SIM\right)\:.
\end{align}
\end{theorem}

\textbf{Theorem \ref{th:OneStep_F_CL}} provides an identity for the successor set of a closed-loop system with the feedback control law described by the set-valued function $\C(x_k)$. For closed-loop successor sets, the input set argument $\U_k$ is omitted and the successor set is instead denoted by $\suc(\R_k,\C)$.

\begin{theorem} \label{th:OneStep_F_CL} \cite[Theorem 3]{Siefert2022} Given a set of states $\R_k\subseteq\Rspace^n$ and closed-loop state-update set $\FSUSCL$, if $\R_{k}\subseteq \text{D}_{\Phi}$ then the\textit{ closed-loop successor set} is given by
\begin{align}
    \label{eq:1stepForward_CL}
    \suc(\R_k,\C) &= \begin{bmatrix}
        \textbf{0} & \id_{n}
        \end{bmatrix}\big(\FSUSCL\cap_{[\id_{n}~\mathbf{0}]} \R_{k}\big) \:.
\end{align}
\end{theorem}

The identities in \eqref{eq:1stepForward_OL}, \eqref{eqn-CLfromOL}, and \eqref{eq:1stepForward_CL} utilize the open-loop state-update set and state-input map. In general, these sets can be complex and difficult to construct in a high-dimensional space using methods that sample or partition the state space. To more efficiently construct these sets, \textbf{Theorem \ref{thm-FunctionCompSets}} utilizes functional decomposition as given in Section~\ref{sec:funcDecomp}, thus only considering unary and binary functions. While \textbf{Theorem~\ref{thm-FunctionCompSets}} addresses nonlinear functions with a single vector argument $x$, it is easily applied to functions with multiple arguments, such as \eqref{eqn-genNLdyn}, by concatenating the arguments into a single vector. \textbf{Theorem \ref{thm-FunctionCompSets}} can be used to construct open-loop state-update sets and state-input maps, as is demonstrated in Section \ref{sect:Examples}.

{
\begin{theorem}
\label{thm-FunctionCompSets}
Consider a general nonlinear function $h(x):\Rspace^n\rightarrow\Rspace^m$ and its decomposition \eqref{eqn-decomp}. Define the set $\Hs$ as
{\small
\begin{align}
    \label{eqn-def-H}
    \Hs \equiv \left\{ \begin{bmatrix}
    w_1\\
    \vdots\\
    w_{n+K}
    \end{bmatrix} \Bigg|\ \begin{array}{c}
         (w_1, w_2, \hdots, w_n) \in \text{D}_{\Hs} ,\\
         w_j = h_j(w_{j1}\{,w_{j2}\})\\
         \forall j\in \{n+1,...,n+K\}
    \end{array} \right\}\:.
\end{align}}%
Given $\text{D}_j \supseteq [e_j]\Hs $ $\forall j=n+1,...,n+K$,\footnote{Sufficiently large $\text{D}_j$ can be found using interval arithmetic.} and
{\small
\begin{align}
    \label{eqn-UnBiFuncSet}
    \Hs_\ell = \left\{ \begin{bmatrix}
    w_{\ell1}\\
    \{w_{\ell2}\}\\
    w_\ell
    \end{bmatrix}\Bigg|\ \begin{array}{cc}
         (w_{\ell1}\{,w_{\ell2}\})\in\text{D}_{\ell1}\{\times \text{D}_{\ell2}\},  \\
         w_\ell = h_\ell(w_{\ell 1}\{,w_{\ell 2}\}) 
    \end{array}\right\}\:,
\end{align}}%
for $\ell \in \{n+1,...,n+K\}$, then $\Hs$ is given by initializing $\Hs_{1:n} = \text{D}_{\Hs}$ and iterating $K$ times through
\begin{align}
    \label{eqn-Hs-Rec}
    \Hs_{1:\ell} &= (\Hs_{1:\ell-1}\times \text{D}_{\ell})\cap_{\begin{bmatrix} 
    e_{\ell1}\\
    \{e_{\ell2}\}\\
    e_\ell
    \end{bmatrix}} \Hs_\ell \:.
\end{align}%
This yields $\Hs_{1:n+K}=\Hs$.

\begin{proof}
From $\Hs_{1:n} = \text{D}_{\Hs}$ and \eqref{eqn-Hs-Rec},
{\small
\begin{align}
\begin{aligned}
\label{eqn-Hcomp-proof}
    &\Hs_{1:n+K} =\\
    \nonumber
    &\left\{ \begin{bmatrix}
        w_1\\
        \vdots\\
        w_{n+K}
    \end{bmatrix} \Bigg| \begin{array}{c}
         (w_1,w_2,...,w_n) \in \text{D}_{\Hs},  \\
         w_j \in \text{D}_j\ \forall j\in\{n+1,...,n+K\}, \\
         w_j = h_j(w_{j1}\{,w_{j2}\})\ \forall j\in\{n+1,...,n+K\}
    \end{array} \right\}\:.
\end{aligned}
\end{align}}

The constraint $w_{n+1}\in \text{D}_{n+1}$ is redundant given that $(w_1,w_2,...,w_n) \in \text{D}_{\Hs}$ and $w_{j} = h_j(w_{j1}\{,w_{j2}\})\ \text{for }j=n+1$, and therefore can be removed. The same is then true $\forall j\in\{n+2,...,n+K\}$, yielding $\Hs$ as defined by \eqref{eqn-def-H}.
\end{proof}
\end{theorem}

\textbf{Theorem~\ref{thm-FunctionCompSets}} addresses compositions of unary and binary functions, however the identity \eqref{eqn-Hs-Rec} is easily extended to enforce relations across many input and output variables, similar to how \eqref{eqn-CLfromOL} couples an arbitrary number of signals between an open-loop plant and a closed-loop controller. \textbf{Corollary \ref{co-affineDecomp}} addresses the special case when the decomposition includes affine functions.
\begin{corollary}
\label{co-affineDecomp}
    Consider a general nonlinear function $h(x):\Rspace^n\rightarrow\Rspace^m$, its decomposition \eqref{eqn-decomp}, the set $\Hs$ defined by \eqref{eqn-def-H}, and the recursion \eqref{eqn-Hs-Rec}. For all $\ell$ where $h_\ell(\cdot)$ is affine, i.e., $h_\ell(w_{\ell1}\{,w_{\ell2}\})=m_{\ell1} w_{\ell1}\{+ m_{\ell2}\ w_{\ell2}\} + b_\ell$ where $m_{\ell1}$, $\{m_{\ell2}\}$, and $b_{\ell}$ are scalars, the set $\Hs$ is equivalently given when \eqref{eqn-Hs-Rec} is replaced by
    \begin{align}
        \Hs_{1:\ell} &= \begin{bmatrix}
            \id_{l-1}\\
            m_{\ell 1}e_{\ell1} \{+m_{\ell 2} e_{\ell2}\}
        \end{bmatrix} \Hs_{1:\ell-1} + \begin{bmatrix}
            \mathbf{0}\\
            1
        \end{bmatrix} b_{\ell} \:.
    \end{align}

    \begin{proof}
        The proof only requires recognizing that $m_{\ell1} w_{\ell1} \{+m_{\ell2}\ w_{\ell2}\} + b_\ell = h_\ell(w_{\ell1}\{,w_{\ell2}\})$ when $h_\ell(\cdot)$ is affine to arrive at \eqref{eqn-Hcomp-proof} and then follows the same procedure as the proof of \textbf{Theorem \ref{thm-FunctionCompSets}}.
    \end{proof}
\end{corollary}

Although \textbf{Corollary \ref{co-affineDecomp}} is written for unary or binary affine functions, it is easily extended to functions with an arbitrary number of arguments, i.e., $h_\ell(w_1,...) = b_\ell + \sum_i m_{\ell i} w_{\ell i}$. Applying \textbf{Corollary~\ref{co-affineDecomp}} reduces the memory complexity of $\Hs$ \eqref{eqn-def-H} when implemented with hybrid zonotopes, as the affine transformation in \eqref{eqn-Hcomp-proof} does not increase the memory complexity of the hybrid zonotope.

Corollary \ref{co-H_inNout} provides an identity to remove intermediate variables that arise from the functional decomposition, and only retains dimensions corresponding to the $n$ arguments and $m$ outputs of $h(x)$. 
\begin{corollary}
\label{co-H_inNout}
    Consider a general nonlinear function $h(x):\Rspace^n\rightarrow\Rspace^m$ and its decomposition \eqref{eqn-decomp}. Given the set $\Hs$ as defined by \eqref{eqn-def-H},
    \begin{align}
        \label{eqn-H_inNout}
        \left\{\begin{bmatrix}
            x\\
            y
        \end{bmatrix}\ \Bigg|\ \begin{array}{c}
             x\in\text{D}_{\Hs}\\
             y= h(x)
        \end{array} \right\} = \begin{bmatrix}
            \id_n & \mathbf{0} & \mathbf{0}\\
            \mathbf{0} & \mathbf{0} & \id_m
        \end{bmatrix} \Hs\:.
    \end{align}%
    \begin{proof}
        By definition of the linear transformation, the right side of \eqref{eqn-H_inNout} yields
        \begin{align}
        \nonumber
    \left\{ \begin{bmatrix}
    w_1\\
    \vdots\\
    w_{n}\\
    w_{n+K-m+1}\\
    \vdots\\
    w_{n+K}
    \end{bmatrix} \Bigg|\ \begin{array}{c}
         (w_1, w_2, \hdots, w_n) \in \text{D}_{\Hs} ,\\
         w_j = h_j(w_{j1}\{,w_{j2}\})\\
         \forall j\in \{n+1,...,n+K\}
    \end{array} \right\} \:.
        \end{align}%
        Substituting $h_j(\cdot)\ \forall\ j\in\{n+1,...,n+K\}$ yields the desired result.
    \end{proof}
\end{corollary}

A fundamental challenge of reachability analysis is that a given set representation and its operations can only achieve efficient computation of \emph{exact} successor sets for a limited class of systems\cite{Gan_RA4SolvableDynSys}. To obtain formal guarantees for broader system classes, \emph{over-approximations} of successor sets are often computed instead \cite{WETZLINGER_AdapReachNonLin}. To this end, \textbf{Corollary \ref{co-SusOA_ReachOA}} extends the previous results by first considering the effect of using over-approximations of sets on the right side of the identities in \eqref{eq:1stepForward_OL}, \eqref{eqn-CLfromOL}, \eqref{eq:1stepForward_CL}, and \eqref{eqn-Hs-Rec}. Then \textbf{Corollary \ref{co-BoundNLFunction}} provides an identity to produce sets that over-approximate nonlinear functions from set-based approximations of those functions and bounds on their approximation error.

\begin{corollary} \label{co-SusOA_ReachOA} For the identities provided by \textbf{Theorems} \textbf{\ref{th:OneStep_F_OL}-\ref{thm-FunctionCompSets}} and \textbf{Corollary \ref{co-H_inNout}}, if any set on the right side is replaced by an over-approximation, then the identity will instead yield an over-approximation of the left side.

\begin{proof}
    Set containment is preserved under linear transformation and generalized intersection.
\end{proof} 
\end{corollary}

For example, if an over-approximation of the open-loop state-update set, given by $\bar{\FSUSOL}$, is used in place of $\FSUSOL$ in \eqref{eq:1stepForward_OL}, then the right side of the equation will yield an over-approximation of the reachable set, $ \suc(\R_k,\U_k)$.

\begin{corollary} \label{co-BoundNLFunction} Given an approximation of a function $\hat{h}_{\ell}(w_{\ell 1}\{,w_{\ell 2}\})\approx h_{\ell}(w_{\ell 1}\{,w_{\ell 2}\})$, its corresponding set $\hat{\Hs}_\ell\approx\Hs_\ell$ from \eqref{eqn-UnBiFuncSet} defined over the same domain $\text{D}_{\hat{\Hs}_\ell}=\text{D}_{\Hs_\ell}$, and an error bound given by the interval $[a,b]$ such that $h_\ell(w_{\ell 1}\{,w_{\ell 2}\}) \in [\hat{h}_\ell(w_{\ell 1}\{,w_{\ell 2}\})+a,\ \hat{h}_\ell(w_{\ell 1}\{,w_{\ell 2}\})+b]$ for all $(w_{\ell 1},\{w_{\ell 2}\})\in\text{D}_{\Hs_\ell}$, an over-approximation of $\Hs_\ell$ is given by
\begin{align}
    \label{eqn-BoundNLFunction}
    \Hs_\ell \subseteq \hat{\Hs}_\ell \oplus \begin{bmatrix}
         0 \\
        \{0\} \\
        1
    \end{bmatrix} \begin{bmatrix}
        a , b
    \end{bmatrix}\:.
\end{align}

\begin{proof}
    Defining the right side of \eqref{eqn-BoundNLFunction} as $\Bar{\Hs}_\ell$, 
    {\footnotesize
    \begin{align}
    \nonumber
        \Bar{\Hs}_\ell = \left\{ 
        \begin{bmatrix}
            \omega_{\ell 1}\\
            \{ \omega_{\ell 2} \} \\
            \omega_\ell
        \end{bmatrix} \Bigg|\ \begin{bmatrix}
             (\omega_{\ell 1},\{\omega_{\ell 2}\})\in\text{D}_{\Hs_\ell},  \\
             \omega_\ell \in \\
             [\hat{h}_\ell(w_{\ell 1},\{w_{\ell 2}\})+a,\ \hat{h}_\ell(w_{\ell 1},\{w_{\ell 2}\})+b] 
        \end{bmatrix} \right\}.
    \end{align}}
\end{proof}
\end{corollary}

\revC{\textbf{Corollary\revM{R3.11, R6.3} \ref{co-BoundNLFunction}} assumes that a bound on the approximation error is known. Generally, for a function $f(x)$ approximated over a domain by an affine approximation $\bar{f}(x)$, error bounds can be posed as the nonlinear programs
\begin{align}
    \label{eq:NLPerr}
    a = \min_{x\in\X}\left( \bar{f}(x)-f(x)\right)\;,\\
    b = \max_{x\in\X}\left( \bar{f}(x)-f(x)\right)\;,
\end{align}}%
where $\X$ is the domain of the approximation. \revM{R3.12}\revC{In general this problem is challenging, though by decomposing functions into unary or binary functions, \eqref{eq:NLPerr} only needs to be solved for each unary and binary nonlinear function of the decomposition, for which $x\in\Rspace^1$ or $\Rspace^2$. Additional properties, such as whether $f(x)$ is continuous, differentiable, convex, or concave, can significantly reduce the complexity of solving \eqref{eq:NLPerr}. Error bounds for affine approximations of some nonlinear functions can be found in \cite[Chapter 3]{wanufelle2007global}, e.g., for $x^2$ a closed-form solution to \eqref{eq:NLPerr} is found by leveraging its convexity and differentiability. In the case of PWA approximations, the process can be repeated for each partition of the domain and the worst-case error bounds should be used.}

Although \textbf{Theorem \ref{thm-FunctionCompSets}} is primarily beneficial for functions with many arguments, Example \ref{ex-unaryDecomp} uses a system with one input to facilitate exposition and visualization. More complex examples of constructing state-update sets of nonlinear systems are considered in Section~\ref{sect:Examples}.

\begin{ex}
\label{ex-unaryDecomp}
A functional decomposition of $x_{k+1}~=~\cos~(~\pi ~\sin~(x_k))$ over a domain $\text{D}_{\Hs}=[-\pi,\pi]$ and its visual representation are given by Table \ref{tab:UnaryDecomp} and Figure \ref{fig:decompUnary}, respectively. Over-approximations $\Bar{\Hs}_2$ and $\Bar{\Hs}_3$ are constructed using SOS approximations and closed-form solution error bounds from \cite{wanufelle2007global}, which are combined using \textbf{Corollary~\ref{co-BoundNLFunction}}. The effect of over-approximation per \textbf{Corollary \ref{co-SusOA_ReachOA}} is demonstrated in Figure \ref{fig:decompUnary}(e)-(h). In Figure \ref{fig:decompUnary}(d) and \ref{fig:decompUnary}(h), \textbf{Corollary \ref{co-H_inNout}} is used as the last step in construction of the state-update set $\Phi$ and its over-approximation $\bar{\Phi}$, which are shown in magenta. 

\renewcommand{\arraystretch}{1.25} 
\begin{table}[h!]
\centering
\caption{Functional decomposition and domain propagation of $x_{k+1}=\cos(\pi \sin(x_k))$ over a domain $\text{D}_{\Hs}=[-\pi,\ \pi]$.}
\begin{tabular}{l|c|c}
$w_j$           & $\Hs_j$                                                   & $\text{D}_j$ \\ \hline \hline
$w_1 = x_k$       &              
 
                                            & $[-\pi\ \pi]$    \\
$w_2$           & $\pi \sin(w_1)$                                           & $[-\pi\ \pi]$    \\
$w_3=x_{k+1}$           & $\cos(w_2)$                                               & $[-1\ 1]$  \\
\end{tabular}
\label{tab:UnaryDecomp}
\end{table}

\begin{figure*}[]
    \centering
     \begin{subfigure}[b]{1.7in}
         \centering
         \includegraphics[width= 1.7 in]{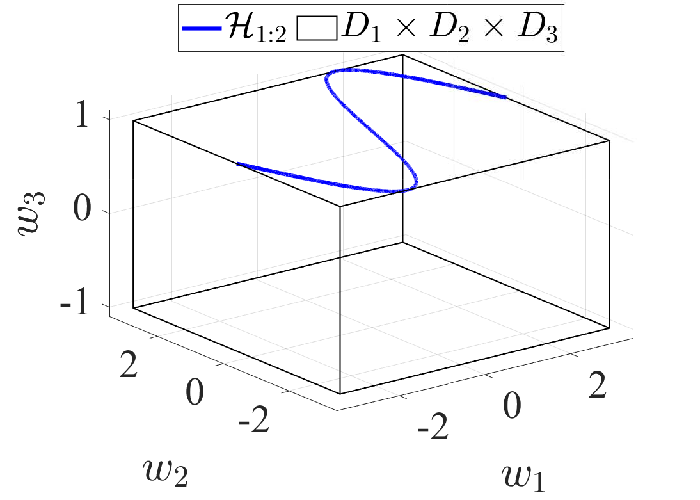}
         \caption{}
         \label{fig:s1_exact}
     \end{subfigure}
     \begin{subfigure}[b]{1.7in}
         \centering
         \includegraphics[width= 1.7 in]{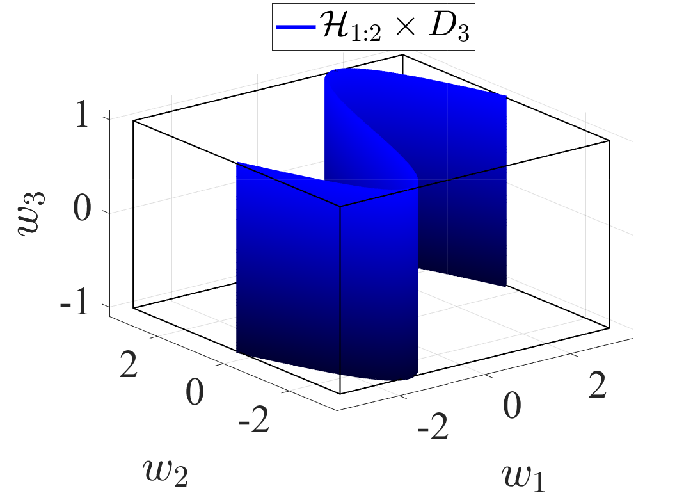}
         \caption{}
         \label{fig:s2_exact}
     \end{subfigure}
     \begin{subfigure}[b]{1.7in}
         \centering
         \includegraphics[width= 1.7 in]{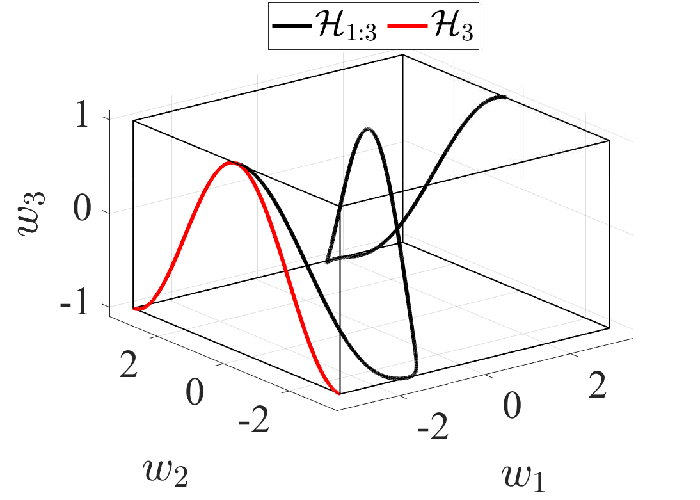}
         \caption{}
         \label{fig:s3_exact}
     \end{subfigure}
     \begin{subfigure}[b]{1.7in}
         \centering
         \includegraphics[width= 1.7 in]{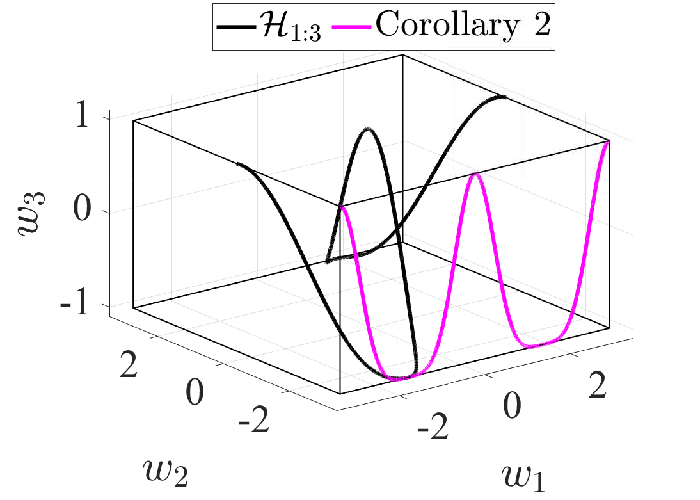}
         \caption{}
         \label{fig:s4_exact}
     \end{subfigure}\\
     \begin{subfigure}[b]{1.7in}
         \centering
         \includegraphics[width= 1.7 in]{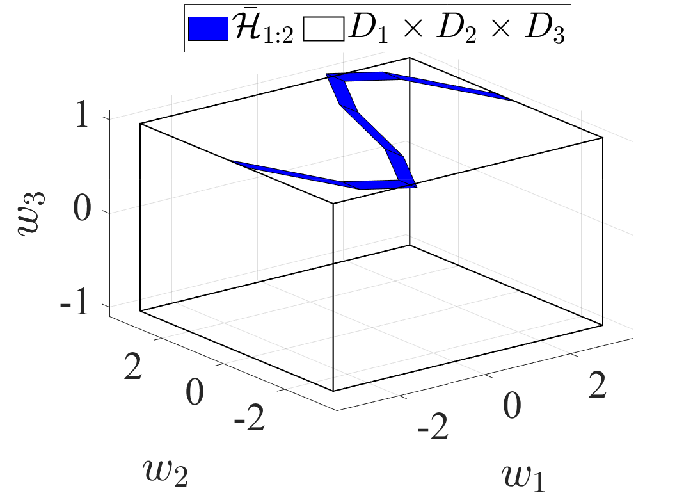}
         \caption{}
         \label{fig:s1_approx}
     \end{subfigure}
     \begin{subfigure}[b]{1.7in}
         \centering
         \includegraphics[width= 1.7 in]{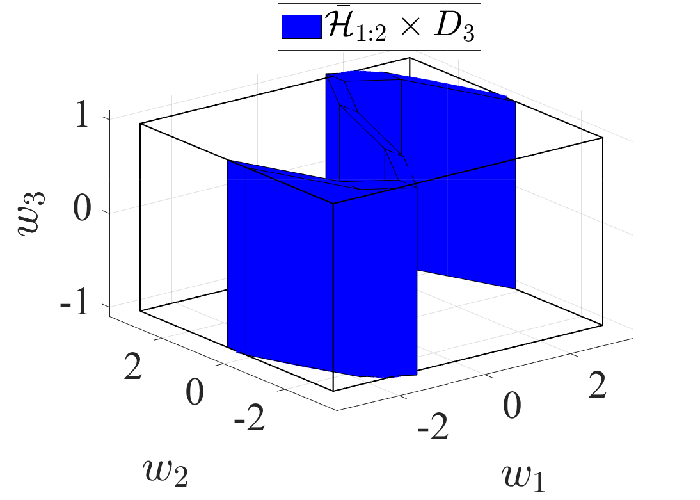}
         \caption{}
         \label{fig:s2_approx}
     \end{subfigure}
     \begin{subfigure}[b]{1.7in}
         \centering
         \includegraphics[width= 1.7 in]{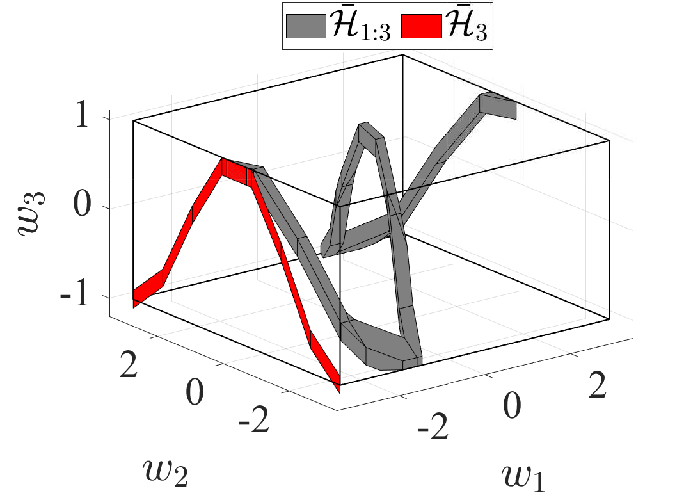}
         \caption{}
         \label{fig:s3_approx}
     \end{subfigure}
     \begin{subfigure}[b]{1.7in}
         \centering
         \includegraphics[width= 1.7 in]{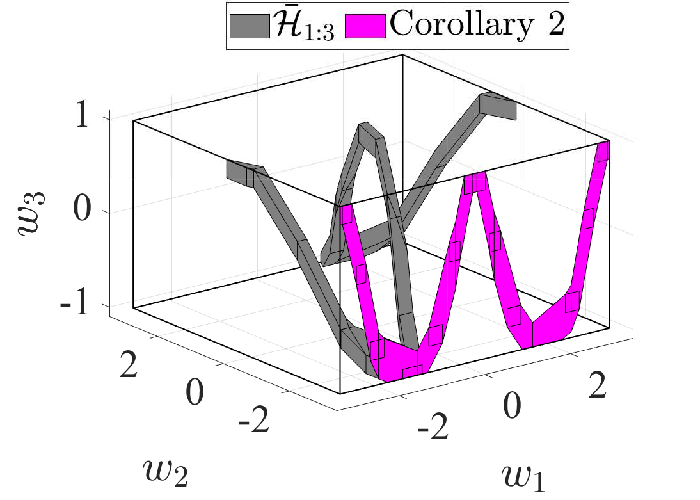}
         \caption{}
         \label{fig:s4_approx}
     \end{subfigure}
     
    \caption{Visual depiction of functional decomposition and \textbf{Theorem \ref{thm-FunctionCompSets}} applied to $x_{k+1}=\cos(\pi \sin(x_k))$ with the decomposition shown in Table~\ref{tab:UnaryDecomp}. Here, $\mathcal{H}_{1:\ell}$ denotes a stage in the recursion of \textbf{Theorem~\ref{thm-FunctionCompSets}}. $\mathcal{H}_{1:1}$ and $\mathcal{H}_{1:1}\times D_2$ are a line and a rectangle, respectively, and are not shown. (a) The first of two recursions of \eqref{eqn-Hs-Rec} yields $\mathcal{H}_{1:2}$ (blue). (b) The first step in the second recursion of \eqref{eqn-Hs-Rec} gives $\mathcal{H}_{1:2}\times D_3$ (also shown in blue). (c) The second step of the second recursion of \eqref{eqn-Hs-Rec} takes a generalized intersection with $\mathcal{H}_3$ (red) and yields $\mathcal{H}_{1:3}$ (black). (d) \textbf{Corollary \ref{co-H_inNout}} eliminates the intermediate variable $w_2$ using a projection and is the last step in producing the state-update set $\Phi$ (magenta). (e-h) Per \textbf{Corollary \ref{co-SusOA_ReachOA}}, over-approximation of the functions used for decomposition results in an over-approximation of the state-update set $\bar{\Phi}\supset \Phi$.}
 \label{fig:decompUnary}
\end{figure*}
\renewcommand{\arraystretch}{1} 
\end{ex}

}

\section{Reachability of Nonlinear Systems Using Hybrid Zonotopes}
\label{sec:HybZonoSUS}


The remainder of this paper assumes that reachable sets and  state-update sets are represented as hybrid zonotopes. Hybrid zonotopes are closed under linear transformations, generalized intersections \cite{Bird_HybZono}, and Cartesian products \cite{BIRDthesis_2022}. The time complexity of the open-loop and closed-loop successor set identities in \eqref{eq:1stepForward_OL} and \eqref{eq:1stepForward_CL}, respectively, is $\mathcal{O}(n)$, as the linear transformations $[\id\ \mathbf{0}]$ under the generalized intersections amount to matrix concatenations. The resulting memory complexity is given by
{\begin{align}
\footnotesize
        \nonumber \revM{R8.4}
        \begin{split}
        \text{\underline{Open}}\\
        n_{g,\text{Suc}} &= n_{g,r}+n_{g,u}+n_{g,\psi},\\
         \nonumber
        n_{b,\text{Suc}} &= n_{b,r}+n_{b,u}+n_{b,\psi},\\
         \nonumber
        n_{c,\text{Suc}} &= n_{c,r}+n_{c,u}+n_{c,\psi}+n,
        \end{split}
        \footnotesize
        \nonumber
        \begin{split}
        \text{\underline{Closed}}\\
        n_{g,\text{Suc}} &= n_{g,r}+n_{g,\phi},\\
         \nonumber
        n_{b,\text{Suc}} &= n_{b,r}+n_{b,\phi},\\
         \nonumber
        n_{c,\text{Suc}} &= n_{c,r}+n_{c,\phi}+n\:.
        \end{split}
\end{align}}%
Reachable sets found by recursion of these identities exhibit linear memory complexity growth with respect to time. 

Because hybrid zonotopes cannot \emph{exactly} represent general nonlinear functions, a trade-off between over-approximation error and complexity arises. The scalability of \textbf{Theorem \ref{thm-FunctionCompSets}} for constructing hybrid zonotopes containing nonlinear functions (for state-update sets, state-input maps, etc.) depends both on the functional decomposition \eqref{eqn-decomp} and the complexity of over-approximating unary and binary functions with hybrid zonotopes. \textbf{Theorem \ref{thm-SOS2HYBZONO}} addresses the latter and additional results for efficient representation of common binary functions are given in Section~\ref{sec:RedMem4Binary}. These are followed in Section~\ref{sec:AvoidCoD} with a discussion of the scalability when a decomposition based on the Kolmogorov Superposition Theorem is used.

\subsection{Converting vertex representation to hybrid zonotope}
 \revC{An identity to convert SOS approximations to hybrid zonotopes was provided in \cite{Siefert2022}. \textbf{Theorem~\ref{thm-SOS2HYBZONO}} provides a more general result for converting a collection of V-rep polytopes into a single hybrid zonotope. Beyond SOS approximations, the result can be applied to represent complex initial or unsafe sets efficiently as a hybrid zonotope.\revM{R8.3}}

\begin{theorem}
\label{thm-SOS2HYBZONO}
{A set $\Ss$ consisting of the union of $N$ V-rep polytopes, $\Ss=\cup_{i=1}^{N}\Pp_i$, with a total of $n_v$ vertices can be represented as a hybrid zonotope with memory complexity 
\begin{align} \label{eqn-SOScomplexity}
    n_{g}=2 n_v,\ n_{b}=N,\ n_{c}=n_{v}+2\:.
\end{align}
\begin{proof}
    Define the vertex matrix $V=[v_1,\dots,v_{n_v}]\in\Rspace^{n\times n_v}$ and construct a corresponding incidence matrix $M\in\Rspace^{n_v \times N}$ with entries $M_{(j,i)}\in\{0,1\} \:\forall \:i,j$, such that 
    \begin{align}
    \nonumber
        \Pp_i=\left\{ V\lambda\ \bigg|\ \lambda_j \in \begin{array}{cc}
             \begin{cases} [0\ 1], & \text{if}\ j\in \{ k\ |\ M_{(k,i)}=1 \}\: \\ \{0\}, & \text{if}\ j\in \{ k\ |\ M_{(k,i)}=0 \} \end{cases}  \\
             \mathbf{1}^T_{n_v} \lambda = 1 
        \end{array} \right\}\:.
    \end{align}%
Define the hybrid zonotope
\begin{equation}\nonumber
    \mathcal{Q}=\frac{1}{2}\left\langle\begin{bmatrix}
        \mathbf{I}_{n_v} \\ \mathbf{0}
    \end{bmatrix},\begin{bmatrix}
        \mathbf{0} \\ \mathbf{I}_{N}
    \end{bmatrix},\begin{bmatrix}
        \mathbf{1}_{n_v} \\ \mathbf{1}_{N}
    \end{bmatrix}, \begin{bmatrix}
        \mathbf{1}_{n_v}^T\\ \mathbf{0}
    \end{bmatrix}, \begin{bmatrix}
        \mathbf{0}\\ \mathbf{1}_{N}^T
    \end{bmatrix},\begin{bmatrix}
        2-n_v \\ 2 -N
    \end{bmatrix}\right\rangle\:,
\end{equation}
and the polyhedron $\mathcal{H}=\{h\in\Rspace^{n_v}~\vert~ h\leq\mathbf{0}\}$, and let
\begin{equation}\label{prop-SOS-eqn-basis}
    \mathcal{D}=\mathcal{Q}\cap_{[\mathbf{I}_{n_v}~-M]}\mathcal{H}\:.
\end{equation}
Then the set $\Ss$ is equivalently given by the hybrid zonotope
\begin{equation}\label{prop-SOS-eqn}
    \Z_{S}=\begin{bmatrix}
        V &\mathbf{0}
    \end{bmatrix}\mathcal{D}\:.
\end{equation}
    By direct application of set operation identities provided in \cite[Section 3.2]{Bird_HybZono}, it can be shown that $\Z_{\Ss}$ has the complexity given by \eqref{eqn-SOScomplexity}. The remainder of the proof shows equivalency of $\Z_\Ss$ and $\Ss$. For any $(\lambda,\delta)\in\mathcal{D}$ there exists some $(\xi^c,\xi^b)\in\mathcal{B}_{\infty}^{n_v}\times\{-1,1\}^{N}$ such that $\mathbf{1}^T_{n_v}\xi^c=2-n_v$, $\mathbf{1}^T_{N}\xi^b=2-N$, $\lambda=0.5\xi^c+0.5\mathbf{1}_{n_v}$, $\delta=0.5\xi^b+0.5\mathbf{1}_{N}$, and $\lambda-M\delta\in\mathcal{H}\implies\lambda\leq M\delta$. 
    Thus $\lambda\in[0,1]^{n_v}$, $\delta\in\{0,1\}^{N}$, $\sum_{i=1}^{n_v}\lambda_i=1$, and $\sum_{i=1}^{N}\delta_i=1$ results in $\delta_i=1\implies\delta_{j\not=i}=0$. 
    Let $\delta_i=1$, then $\lambda\leq M\delta$ enforces $\lambda_j\in[0,1]\: \forall\ j\in\{k\ |\ M_{(k,i)}=1\}$ and $\lambda_j=0\ \forall\ j\in\{k\ |\ M_{(k,i)}=0\}$. Therefore given any $z\in\mathcal{Z}_{S}$ corresponding to $\delta_i=1$, 
    \begin{align}
    \label{eqn-zinZs}
        z=\sum \lambda_j v_j\:\forall\ j\in\{k\ |\ M_{(k,i)}=1\}\:,
    \end{align}%
    thus $z\in\Pp_i\subseteq\Ss$ and $\mathcal{Z}_{S}\subseteq~\Ss$. 
    
    Conversely, given any $x\in\Ss$, $\exists\ i$ such that $x\in\Pp_i=\sum \lambda_j v_j\:\forall\ j\in\{k\ |\ M_{(k,i)}=1\}$. This is equivalent to \eqref{eqn-zinZs}, therefore $x\in\Z_\Ss$, $\Ss\subseteq \Z_\Ss$, and $\Z_\Ss = \Ss$.
\end{proof}}
\end{theorem}

\revC{The \revM{R3.13} computational complexity of \textbf{Theorem \ref{thm-SOS2HYBZONO}} is ${\mathcal{O}(n(n_v+N)^2)}$}. The pre-existing approach for converting a collection of V-rep polytopes to a single hybrid zonotope would be to convert each polytope from V-rep to H-rep to HCG-rep and then take iterative unions using the method given in \cite{Bird_HybZonoUnionComp}. 
However, this would involve greater computational complexity, as conversion between V-rep and H-rep alone has worst-case exponential computational complexity \cite{scott2016constrained}, and would produce a set with memory complexity that scales quadratically with the number of polytopes. Furthermore \textbf{Theorem \ref{thm-SOS2HYBZONO}} allows for the intuitive representation of sets as hybrid zonotopes through the use of the vertex matrix and incidence matrix, as demonstrated in Example \ref{ex-Triangle}.
\begin{ex}
\label{ex-Triangle}
Consider a vertex matrix $V=[v_1,v_2,v_3]$ consisting of the vertices of a triangle. The set of vertices, the set of points along the edges of the triangle, and the convex hull of the vertices can be found as a hybrid zonotope using \textbf{Theorem \ref{thm-SOS2HYBZONO}} and the respective incidence matrices
\begin{align}
    \nonumber
    M_{vertices} = \id_3\:,\
    M_{edges} = \begin{bmatrix}
        1 & 0 & 1\\
        1 & 1 & 0\\
        0 & 1 & 1
    \end{bmatrix}\:,\
    M_\Delta = \begin{bmatrix}
        1\\ 
        1\\ 
        1
    \end{bmatrix}\:.
\end{align}%
\end{ex}
 Example \ref{ex-sinxHZ} demonstrates the use of \textbf{Theorem \ref{thm-SOS2HYBZONO}} to represent an SOS approximation of $\sin(x)$ as a hybrid zonotope. 
\begin{ex} \label{ex-sinxHZ}
Figure \ref{fig-SINXHZ} shows $y=\sin(x)$ (green) for $x\in[-4,4]$ and an SOS approximation (red) with the vertex and incidence matrices given in Example~\ref{ex-sinSOS}. The SOS approximation $\Z_{S}$ is represented as a hybrid zonotope using \textbf{Theorem \ref{thm-SOS2HYBZONO}}. An envelope $\Bar{\Z}_{\sin(x)} \supset \left\{ (x,\sin(x))\ |\ x\in[-4,4] \right\}$, shown in blue, is calculated using \textbf{Corollary \ref{co-BoundNLFunction}} and rigorous error bounds for SOS approximations given by \cite[Chapter 3]{wanufelle2007global}. 
\end{ex}
\begin{figure*}[]
    \centering
    \includegraphics[width=0.6\linewidth]{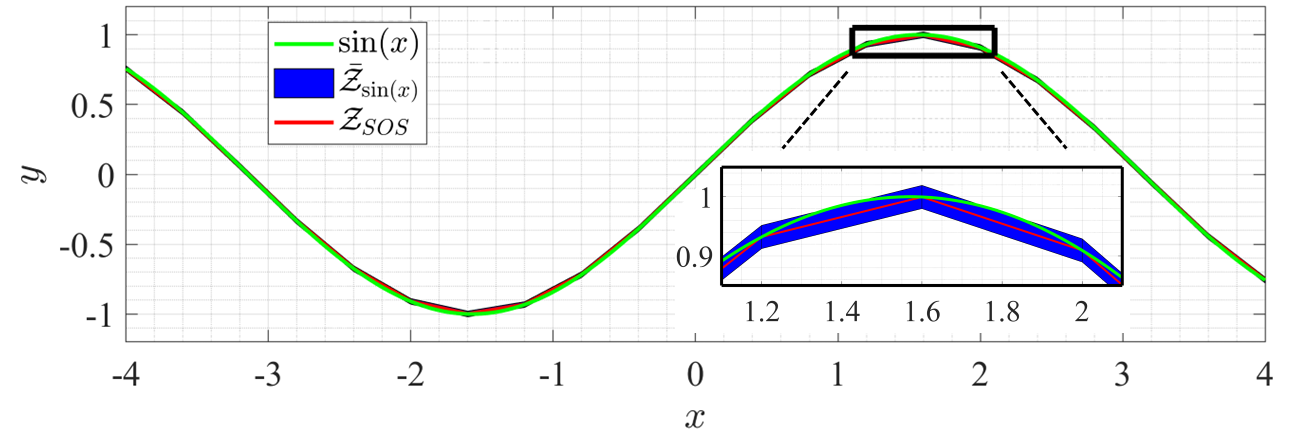}
    \caption{A sinusoid (green) is approximated using an SOS approximation for $x\in[-4,4]$ and represented as a hybrid zonotope (red). Using formal bounds for SOS approximation error, the SOS approximation is bloated in the output dimension to create an enclosure of a sine wave for $x\in[-4,4]$, which is also represented as a hybrid zonotope (blue).}
    \label{fig-SINXHZ}
\end{figure*}

\subsection{Common binary functions}
\label{sec:RedMem4Binary}

This section demonstrates how the memory complexity of hybrid zonotopes approximating graphs of several common binary functions can be reduced as compared to sampling in both arguments of the functions. This is achieved using functional decompositions composed exclusively of unary nonlinear functions and affine functions, leveraging separable functions \cite[Chapter 7.3]{williams2013model} and extensions presented in \cite[Section 4.2]{SZUCS_2012_OptimalPWA}. When nonlinear functions within functional decompositions are strictly unary, SOS approximations can be constructed by sampling one-dimensional spaces. For example, in the case of $xy$, $\frac{x}{y}$, and $x^y$, where $x,y\in\Rspace$, this reduces the task from sampling a two-dimensional space to sampling a one-dimensional space $2$, $3$, and $4$ times respectively. The decompositions for each of these expressions are shown in Table~\ref{tab:CommonBinaryFD} and graphs of the decompositions can be constructed using methods presented in Section~\ref{sect:SUS_reach}.

\renewcommand{\arraystretch}{1.4} 
\begin{table}[]
    \centering
    \caption{Functional decompositions of $xy$, $\frac{x}{y}$, and $x^y$ using 2, 3, and 4 unary nonlinear functions, respectively, with additional affine functions. Equivalency of each binary function with the highest indexed variable can be shown using substitution.}
    \begin{tabular}{c||c|c|c}
         & $xy$ & $\frac{x}{y}$ & $x^{y}$ \\ \hline \hline
         $w_1$ & $x$ & $x$ & $x$ \\ \hline
         $w_2$ & $y$ & $y$ & $y$ \\ \hline
         $w_3$ & $w_1+w_2$ & $\frac{1}{w_2}$ & $\ln w_1$ \\ \hline
         $w_4$ & $w_1-w_2$ & $w_1 + w_3$ & $w_1 + w_3$ \\ \hline
         $w_5$ & $w_3^2$ & $w_1 - w_3$ & $w_1 - w_3$ \\ \hline
         $w_6$ & $w_4^2$ & $w_4^2$ & $w_4^2$ \\ \hline
         $w_7$ & $\frac{1}{4}(w_5-w_6)$ & $w_5^2$ & $w_5^2$ \\ \hline
         $w_8$ & - & $\frac{1}{4}(w_6-w_7)$ & $\frac{1}{4}(w_6-w_7)$ \\ \hline
         $w_9$ & - & - & $e^{w_8}$ \\ 
    \end{tabular}
    \label{tab:CommonBinaryFD}
\end{table}
\renewcommand{\arraystretch}{1} 

\begin{ex}
    \label{ex-BLreduction}
    To demonstrate the scalability advantage of constructing graphs of binary functions using functional decompositions with only affine functions and unary nonlinear functions, consider three methods for building a hybrid zonotope approximation of the bilinear function $xy$. These methods are: (M1) uniformly sampling the two-dimensional input space and generating a hybrid zonotope using \textbf{Theorem~\ref{thm-SOS2HYBZONO}}, (M2) decomposing the function according to Table~\ref{tab:CommonBinaryFD}, sampling the $w_5(w_3)$ and $w_6(w_4)$ unary functions, generating hybrid zonotope approximations of the quadratic functions using \textbf{Theorem~\ref{thm-SOS2HYBZONO}}, and constructing a graph of the bilinear function using methods from Section~\ref{sect:SUS_reach}, and (M3) uniformly sampling in one dimension and generating a hybrid zonotope using \textbf{Theorem~\ref{thm-SOS2HYBZONO}}. The intuition of M3 is that for a fixed value of $x$, $xy$ is linear with respect to the $y$ dimension. Thus it is possible to to obtain tighter over-approximations of $xy$ by only increasing sampling of the $x$ dimension (or the $y$ dimension). 
    
    The domain of interest is specified as $(x,y)\in[-1, 1]^2\rightarrow(w_3,w_4)\in[-2,2]^2$. For M1 and M2, $n_{x,\text{M}1}=n_{y,\text{M}1}$ and $n_{w_3,\text{M}2}=n_{w_4,\text{M}2}$. The number of samples in the $w_3$ and $w_4$ spaces, $n_{w_3,\text{M}2}$, is chosen as $n_{w_3,\text{M}2} = \lceil \sqrt{2}n_{x,\text{M}1} \rceil$ to ensure that the spacing of the approximation for M2 partitions the domain into squares with a side length at least as small as those using M1. To compare the scalability of M3 with M2, the sampling for M3 is chosen to result in the same number of binary factors as M2 by setting $n_{x,\text{M}3}=2n_{5,\text{M}2}$ and $n_{y,\text{M}3}=2$.

    Table \ref{tab-bilinearComplexityCompare} compares the complexity of the three methods, demonstrating how decomposing binary functions into affine functions and unary nonlinear functions enables more scalable over-approximations of the bilinear function as the sampling is increased and over-approximations become more accurate. Figure \ref{fig:BL-m1vm2} plots over-approximations of the bilinear function using all three methods, corresponding to Case $(6,17,34)$ in Table \ref{tab-bilinearComplexityCompare}. The partitioning in Figure \ref{fig:BL-m1vm2}(a) and Figure \ref{fig:BL-m1vm2}(c) corresponds to sampling in the $(x,y)$ space while the partitioning in Figure \ref{fig:BL-m1vm2}(b) corresponds to sampling in the $(w_3,w_4)$ space.

    \begin{figure*}[]
        \centering
        \begin{subfigure}[b]{2.3in}
         \centering
         \includegraphics[width=2.3in]{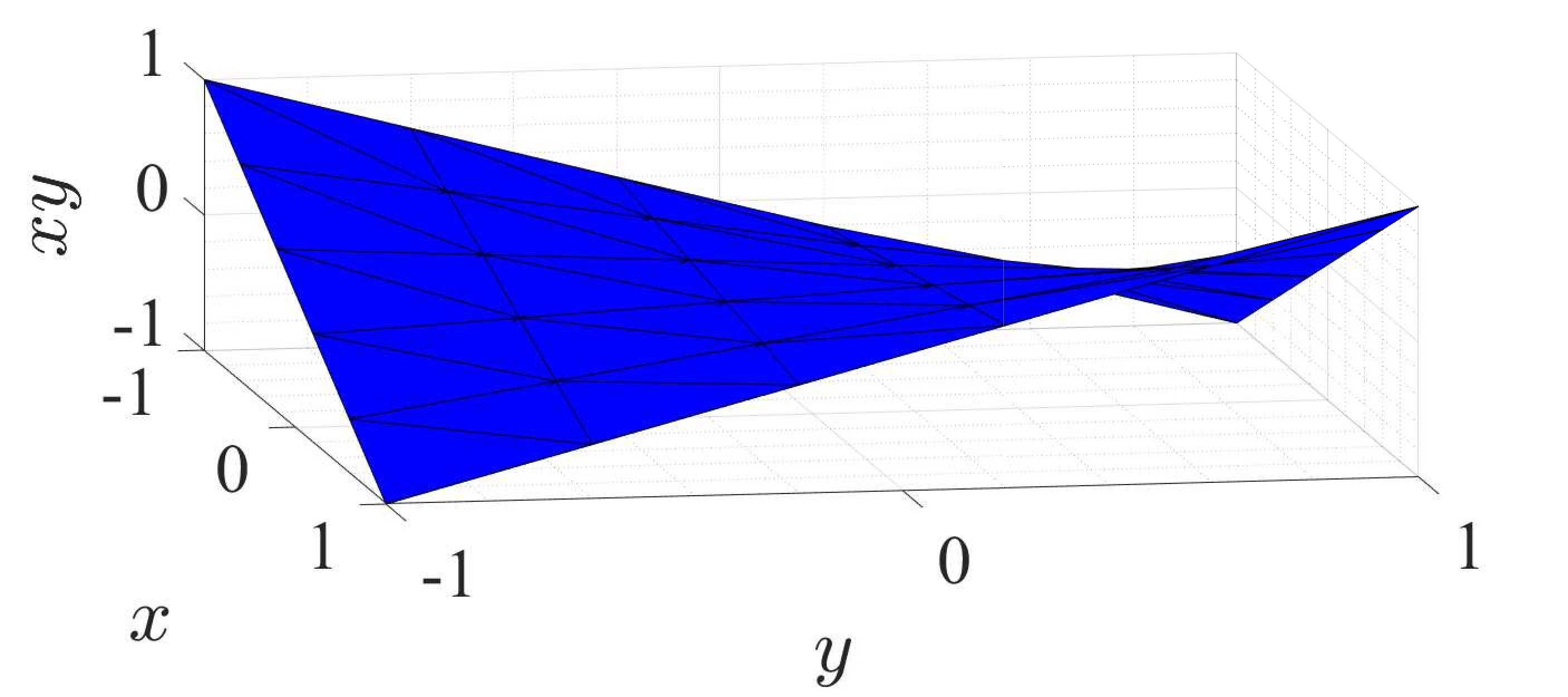}
         \caption{M1: $n_x=6$}
         \label{fig:BL-m1}
     \end{subfigure}
    \begin{subfigure}[b]{2.3in}
         \centering
         \includegraphics[width=2.3in]{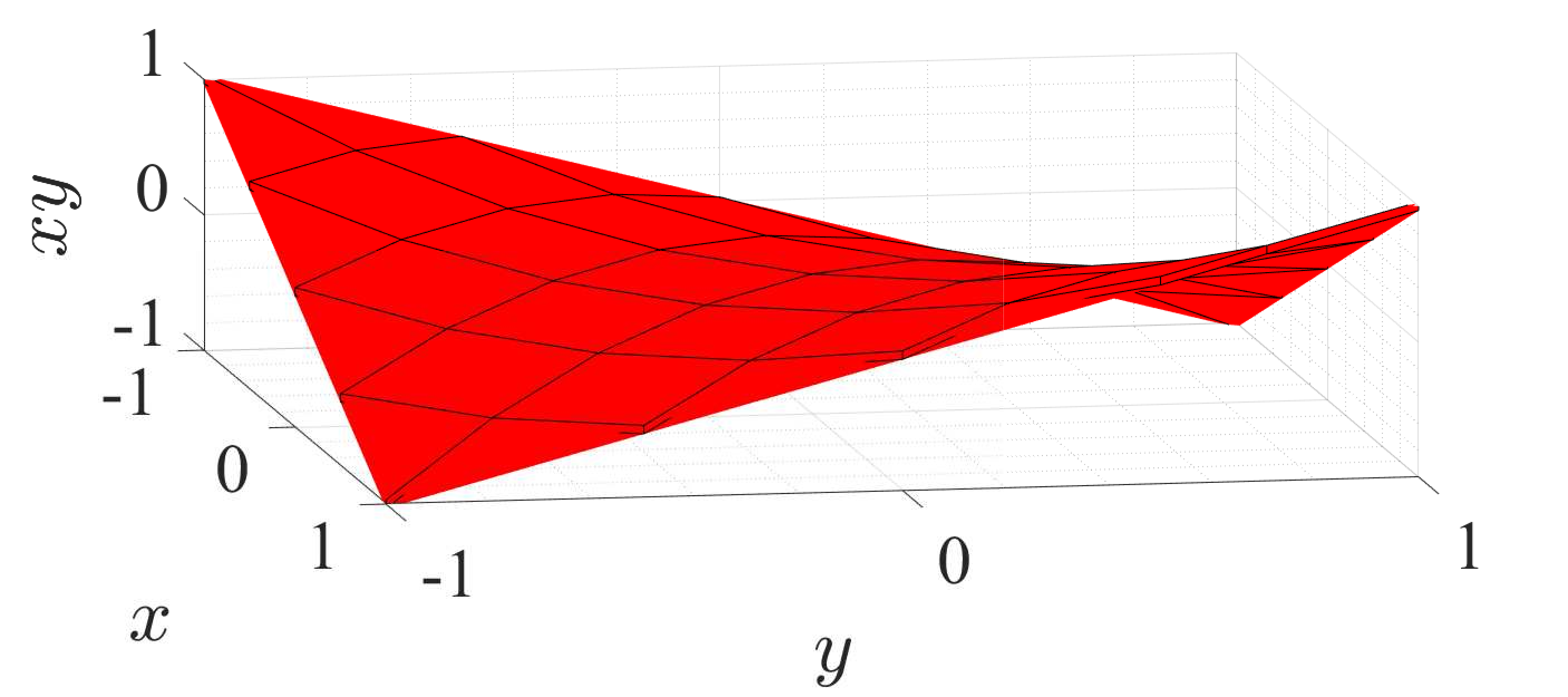}
         \caption{M2: $n_3=17$}
         \label{fig:BL-m2}
     \end{subfigure}
     \begin{subfigure}[b]{2.3in}
         \centering
         \includegraphics[width=2.3in]{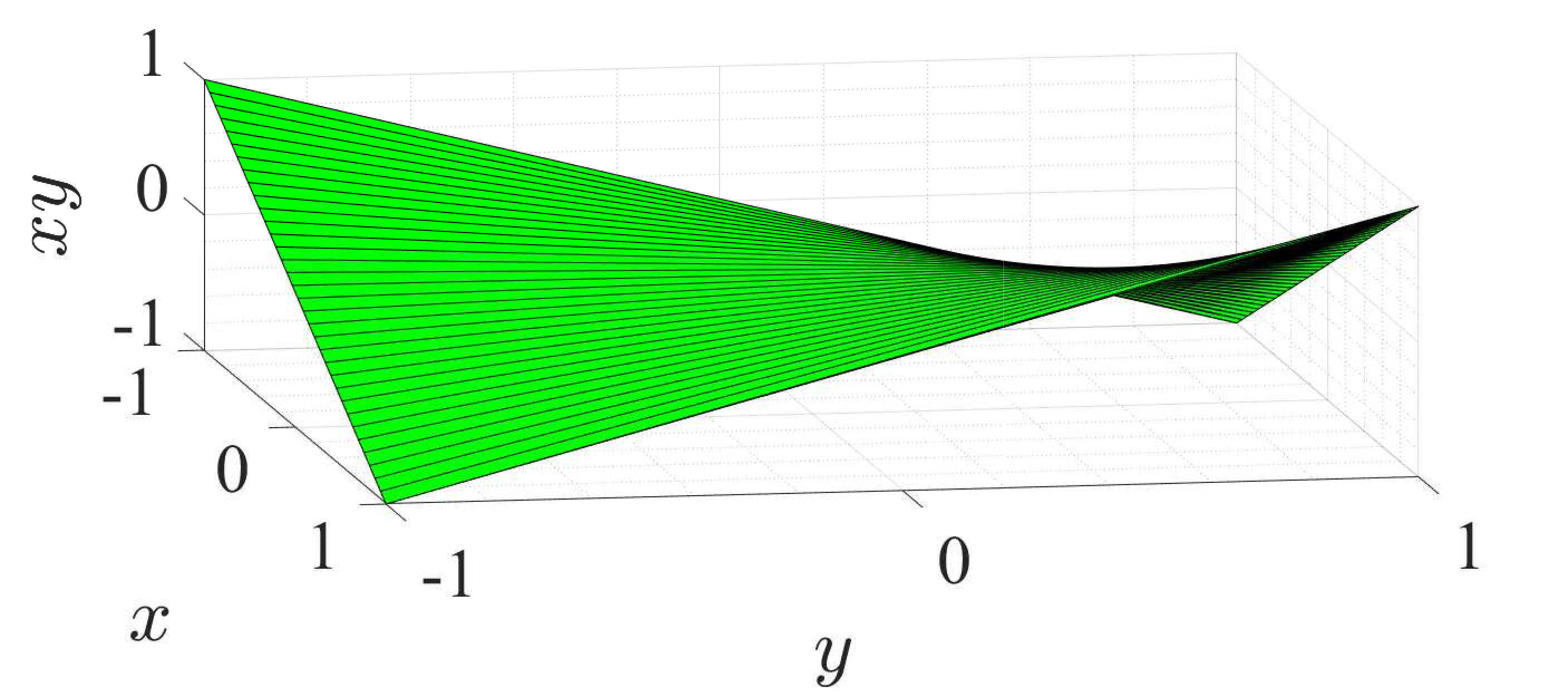}
         \caption{M3: $n_x=34,\ n_y=2$}
         \label{fig:BL-m3}
     \end{subfigure}
      \caption{Comparison of three methods for over-approximating $\{\begin{bmatrix}
          x & y & xy
      \end{bmatrix}^T)\ |\ (x,y)\in[-1\ 1]^2\}$.}
      \label{fig:BL-m1vm2}
    \end{figure*}

    \begin{table}[]
        \centering
        \caption{Complexity comparison of three methods for approximating the bilinear function $f(x,y)=xy$. The leftmost column denotes ($n_x$ for M1,\, $n_3$ for M2,\, $n_x$ for M3) and these values are selected for parity in their approximation accuracy.}
        \begin{tabular}{c|c|c|c}
            $\begin{array}{c}
                 \text{Case }  
            \end{array}$   &  $\begin{array}{c}
                 \text{M1}  \\
                 n_x=n_y
            \end{array}$ & $\begin{array}{cc}
                 \text{M2}  \\
                 n_3=n_4 
            \end{array}$ & $\begin{array}{c}
                 \text{M3}  \\
                 n_x\ (n_y = 2) 
            \end{array}$ \\ \hline \hline
            (3,9,18)  &  $\begin{array}{l}
                 n_g = 19  \\
                 n_b = 5  \\
                 n_c = 11  
            \end{array}$ &   $\begin{array}{l}
                 n_g = 24  \\
                 n_b = 10  \\
                 n_c = 14  
            \end{array}$ & $\begin{array}{l}
                 n_g = 41  \\
                 n_b = 10  \\
                 n_c = 22  
            \end{array}$ \\ \hline
            (6,17,34)  &   $\begin{array}{l}
                 n_g = 73  \\
                 n_b = 26  \\
                 n_c = 38  
            \end{array}$ & $\begin{array}{l}
                 n_g = 40  \\
                 n_b = 18  \\
                 n_c = 22  
            \end{array}$ & $\begin{array}{l}
                 n_g = 73  \\
                 n_b = 18  \\
                 n_c = 38 
            \end{array}$ \\ \hline
            (9,26,52)  &   $\begin{array}{l}
                 n_g = 163  \\
                 n_b = 65  \\
                 n_c = 83  
            \end{array}$ & $\begin{array}{l}
                 n_g = 56  \\
                 n_b = 26  \\
                 n_c = 30  
            \end{array}$ & $\begin{array}{l}
                 n_g =  105 \\
                 n_b =  26 \\
                 n_c =  54
            \end{array}$\\ \hline
            (12,34,68)  &   $\begin{array}{l}
                 n_g = 289  \\
                 n_b = 122  \\
                 n_c = 146  
            \end{array}$ & $\begin{array}{l}
                 n_g =  72 \\
                 n_b =  34 \\
                 n_c =  38 
            \end{array}$ & $\begin{array}{l}
                 n_g =  137 \\
                 n_b =  34 \\
                 n_c =  70
            \end{array}$
        \end{tabular}
        \label{tab-bilinearComplexityCompare}
    \end{table}
\end{ex}

\subsection{Avoiding the curse of dimensionality} 
\label{sec:AvoidCoD}
This section provides a theoretical bound on the memory complexity of the open-loop state-update set $\Psi$ and state-input map $\Theta$ (together these bound the complexity of the closed-loop state-update set $\Phi$ via \textbf{Theorem \ref{thm-CLSUSfromOL}}) to quantify the scalability of the proposed methods with respect to the state dimension and number of vertices used to approximate nonlinear functions. 

Consider a general nonlinear function $h(x):\Rspace^n\rightarrow\Rspace^m$. We assume that the functional decomposition consists of $n$ intermediate variables corresponding to the dimensions of $x$, $K_{aff}$ intermediate variables corresponding to multivariate affine functions, and $K_{NL}$ intermediate variables corresponding to nonlinear functions that are unary or binary. We refer to the set of indices such that $h_\ell(\cdot)$ is nonlinear as $\mathcal{N}$. \revM{R3.5, R5.5}\revC{Assuming $\text{D}_\Hs$ and $\text{D}_\ell$ \eqref{eqn-Hs-Rec} are intervals, the memory complexity of $\Hs$ as constructed by \textbf{Theorem \ref{thm-FunctionCompSets}} and \textbf{Corollary \ref{co-affineDecomp}} in terms of the complexities of $H_\ell ,\ \forall \ell$ such that $h_\ell(\cdot)$ is nonlinear, is given by}

{\small \revC{
\begin{align}
\revM{R3.5}
    n_{g,\Hs} &= n+K_{NL} + \sum_{\ell \in \mathcal{N}} n_{g,\Hs_\ell} \:, \\
    n_{b,\Hs} &= \sum_{\ell \in \mathcal{N}} n_{b,\Hs_\ell} \:, \\
    n_{c,\Hs} &= \sum_{\ell \in \mathcal{N}} \left(n_{c,\Hs_\ell}+n_\ell \right),\quad 
    n_\ell =
    \begin{cases}
         2 & \text{if } h_\ell(\cdot) \text{ is unary} \:, \\
         3 & \text{if } h_\ell(\cdot) \text{ is binary} \:. 
    \end{cases} 
\end{align}}}%
\revC{The computational complexity of constructing $\Hs$ using \textbf{Theorem \ref{thm-FunctionCompSets}} and \textbf{Corollary \ref{co-affineDecomp}} is \revM{R3.15}dominated by the $K_{NL}$ generalized intersection recursions of \eqref{eqn-Hs-Rec}. Each generalized intersection has computational complexity $\mathcal{O}(\ell (n_g + n_b))$, where $n_g$ and $n_b$ grow with the complexity of $\Hs_\ell$.}

For a given system, the decomposition plays a critical role in the memory complexity of the resulting set $\Hs$ defined in \eqref{eqn-def-H}. Thus, quantifying the scalability of the proposed approaches for constructing graphs of functions with respect to the state dimension is challenging, as it is expected that $K_{NL}$ will grow with $n$. \revC{To provide a theoretical bound for the proposed method, let us assume that a decomposition based on the Kolmogorov Superposition Theorem \cite{KolmogorovTheoremRelevant} is performed for each $h_i(x),\ \forall\ i \in \{1,...,m\}$. Then the number of unary nonlinear decomposition functions is given by $K_{NL}=2mn^2 + mn$ and no binary nonlinear decomposition functions are required. Assuming that each unary nonlinear decomposition function is approximated using an SOS approximation with $n_v$ vertices, the resulting complexity of $\Hs$ is given by}
\begin{align}
\nonumber
    n_{g,\Hs} &= n + (2mn^2 + mn)(2n_v+1) \:, \\
    \label{eqn-Hcomp-Kolmog}
    n_{b,\Hs} &= (2mn^2 + mn)(n_v-1) \:, \\
\nonumber
    n_{c,\Hs} &= 8mn^2 + 4mn \:. 
\end{align}
It is clear from \eqref{eqn-Hcomp-Kolmog} that the memory complexity of $\Hs$ scales as a polynomial with $n$, avoiding the exponential growth associated with the curse of dimensionality. In comparison, any method that sampled the $n$-dimensional space to generate an approximation would scale as $n_v^n$, not including any additional complexity incurred to generate $\Hs$ from those samples.

\section{Numerical Examples}
\label{sect:Examples}

Results in this section were generated with MATLAB on a desktop computer with a 3.0 GHz Intel i7 processor and 16 GB of RAM. Reachable sets were plotted using techniques from \cite{Bird_HybZono} and \cite{BIRDthesis_2022}.

\subsection{Single pendulum controlled by a neural network} \label{sec:SinglePendulmEx_OURS}
This numerical example demonstrates reachability analysis of a nonlinear inverted pendulum in closed loop with a neural network controller trained to mimic Nonlinear Model Predictive Control (NMPC). This is accomplished by constructing an over-approximation of the open-loop state-update set $\Psi$ (\ref{sec-ex1-psi}) and a state-input map for the neural network controller $\Theta$ (\ref{sec-ex1-theta}), which are then combined to form an over-approximation of the closed-loop state-update set $\Phi$ (\ref{sec-ex1-phi}) and used for reachability analysis (\ref{sec-ex1-reach}). 

Consider the dynamics of an inverted pendulum given by \eqref{eqn-PendulumContinuous}
with gravity $g=10$, length $l=1$, mass $m=1$, and moment of inertia $I=ml^2=1$. The continuous-time nonlinear dynamics are discretized with time step $h=0.1$ using a $2^{nd}$-order Taylor polynomial $\mathcal{T}_2(x_k)$ \eqref{eqn-T2}, and over-approximated as
{
\begin{align} \label{eq:Taylorpoly_L}
    x_{k+1} &\in \mathcal{T}_2(x_k) \oplus \mathcal{L}\:,
\end{align}}%
where $\mathcal{L} $ is constructed using the Taylor inequality to bound the error due to truncating higher-order terms. The input torque is controlled in discrete time with a zero-order hold and bounded as $u_k\in[-20, 20]\ \forall k$.
\subsubsection{Construction of open-loop state-update set} \label{sec-ex1-psi}
A functional decomposition of $\mathcal{T}_2(x_k)$ is performed and, for a chosen $\text{D} (\Hs)=\text{D}_1 \times \text{D}_2 \times \text{D}_3$, domain propagation is performed to determine $\text{D}_j\ \forall j\in\{4,5,6\}$. This is shown in Table~\ref{tab:PendulumDecomp}. This choice of the domain also allows us to construct $\mathcal{L}$ in \eqref{eq:Taylorpoly_L} as $\mathcal{L} = \begin{bmatrix} -0.02, 0.02 \end{bmatrix}\times\begin{bmatrix} -0.26, 0.26 \end{bmatrix}$. The sets $\Bar{\Hs}_\ell\ \forall \ell \in\{4,5,6\}$ are constructed using \textbf{Theorem \ref{thm-SOS2HYBZONO}}, \textbf{Corollary \ref{co-BoundNLFunction}} and error bounds given for SOS approximations in \cite{leyffer2008branch}. The set $\Bar{\Hs}$ using \textbf{Theorem \ref{thm-FunctionCompSets}} with a Minkowski summation to account for the Taylor remainder $\mathcal{L}$ yields an over-approximation of the open loop state-update set,
\begin{align}
    \Psi \subset \Bar{\Psi} = \Bar{\Hs} \oplus \begin{bmatrix}
        0 & 0\\
        0 & 0\\
        0 & 0\\
        1 & 0\\
        0 & 1
    \end{bmatrix}\mathcal{L}\:.
\end{align}%
 A projection of $\Bar{\Psi}$ is shown in Figure \ref{fig:SUS_SinglePendulumOurs}(a). The thickness of the set in the $x_{2,k}$ dimension is primarily due to the open-loop state-update set capturing variability in the input $u_k\in[-20, 20]$, though some of this is also a result of the Taylor remainder $\mathcal{L}$ and over-approximation of nonlinear functions, with $\Bar{\Hs}_\ell\supset{\Hs}_\ell\ \forall \ell\in\{4,5,6\}$. 
 
Computing $\Bar{\Hs}_\ell\ \forall \ell\in\{4,5,6\}$ required 0.02 seconds, and  application of exact complexity reduction techniques from~\cite{Bird_HybZono,BIRDthesis_2022} required an additional 2 seconds. From this, $\bar{\Psi}$ with memory complexity given in Table~\ref{tab:ComplexitySets_Ex1}, was computed in 8 milliseconds using \textbf{Theorem \ref{thm-FunctionCompSets}} and \textbf{Corollary \ref{co-affineDecomp}}.

\begin{figure}
    \centering
    \begin{subfigure}[b]{2.75in}
         \centering
         \includegraphics[width=2.75in]{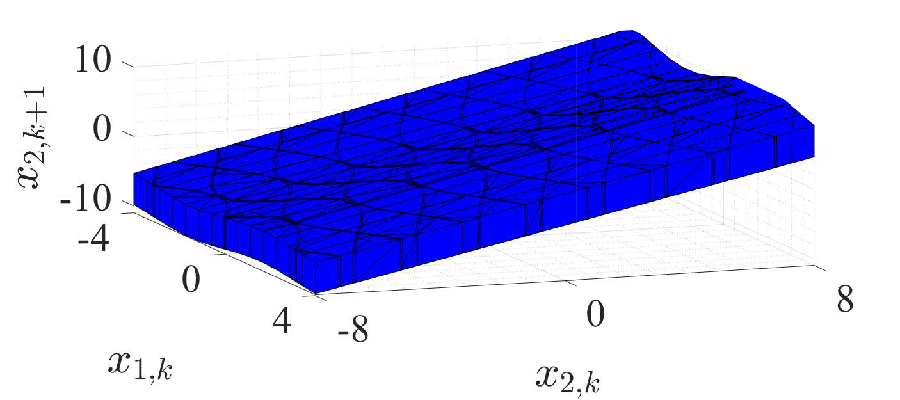}
         \caption{Projection of $\bar{\FSUSOL}$}
         \label{fig:PSI124}
     \end{subfigure}\\
    \begin{subfigure}[b]{2.75in}
         \centering
         \includegraphics[width=2.75in]{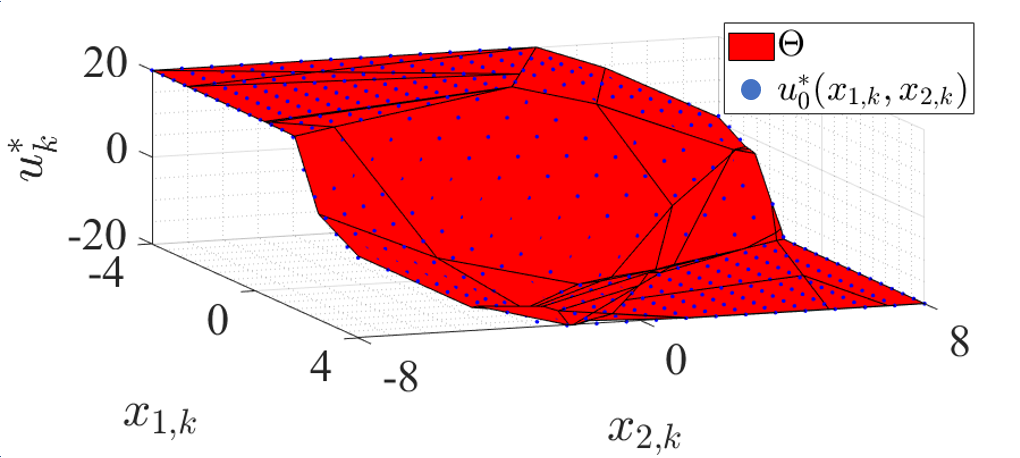}
         \caption{$\SIM$}
         \label{fig:SIM}
     \end{subfigure}\\
    \begin{subfigure}[b]{2.75in}
         \centering
         \includegraphics[width=2.75in]{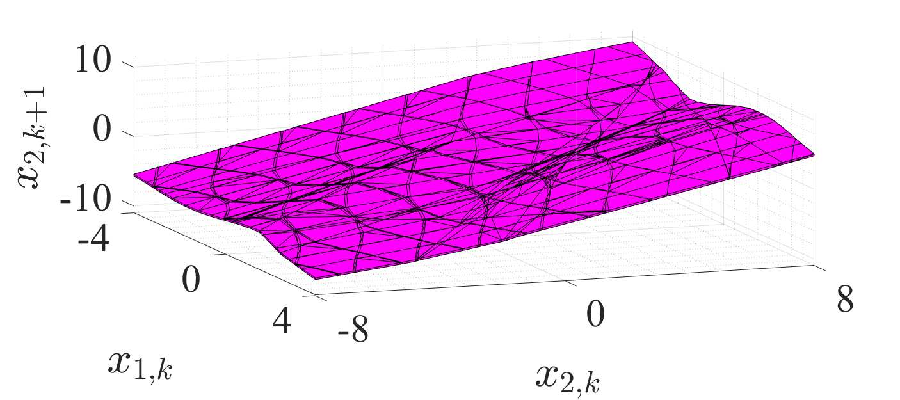}
         \caption{Projection of $\bar{\FSUSCL}$}
         \label{fig:PHI124}
     \end{subfigure}
    \caption{(a)~{Projection of over-approximated open-loop state-update set $\Bar{\Psi}$ bounding dynamics of a pendulum at discrete time steps}. (b)~{State-input map $\SIM$ of a neural network trained to mimic NMPC}. (c)~Projection of over-approximated closed-loop state-update set found using \textbf{Theorem \ref{thm-CLSUSfromOL}} by coupling $\Bar{\Psi}$ and $\Theta$.}
    \label{fig:SUS_SinglePendulumOurs}
    \vspace*{-\baselineskip}
\end{figure}
\renewcommand{\arraystretch}{1.25} 

\begin{table}[htb!]
    \caption{Memory complexity of the open-loop state-update set, state-input map, and closed-loop state-update set.}
    \centering
    \begin{tabular}{c|c|c|c}
        Set & $n_g$ & $n_b$ & $n_c$  \\ \hline \hline
        $\Bar{\Psi}$ & $113$ & $45$ & $63$\\ \hline
        $\Theta$ & $76$ & $20$ & $56$\\ \hline
        $\Bar{\Phi}$ & $184$ & $63$ & $117$
    \end{tabular}
    \label{tab:ComplexitySets_Ex1}
\end{table}
\renewcommand{\arraystretch}{1} 

\subsubsection{Construction of state-input map}
\label{sec-ex1-theta}
To train the neural network controller, an NLMPC is formulated as
\begin{align}
    \label{eqn-NLMPCform}
    \min_{u(\cdot)}&\quad \sum_{k=1}^{10} x_k^T \begin{bmatrix}
        100 & 0\\
        0 & 1
    \end{bmatrix} x_k + u_{k-1}^T u_{k-1}\\
    \nonumber
    \text{s.t.}&\quad \text{trapezoidal discretization of \eqref{eqn-PendulumContinuous} holds}\:,\\
    \nonumber
    &\quad u_k\in[-20, 20] \: \forall k\in\{0,...,9\}\:. 
\end{align}%
The solution of \eqref{eqn-NLMPCform} is found using the MATLAB Model Predictive Control Toolbox \cite{MATLAB:2010} for 400 uniformly sampled initial conditions. The sampled initial conditions and the first optimal input of the solution trajectory $u_0^*$ are then used as input-output pairs to train a neural network with 2 hidden layers, each with 10 nodes and Rectified Linear Unit (ReLU) activation functions, using the MATLAB Deep Learning Toolbox. A functional decomposition of the neural network with saturated output to obey torque constraints is performed and \textbf{Theorem \ref{thm-FunctionCompSets}} is used to generate the state-input map $\Theta$ in a similar fashion as the over-approximation of the open-loop state-update set $\Psi$. This decomposition is omitted for brevity, however we note that the state-input map can be represented \textit{exactly} in this case, as the only nonlinear functions involved are the ReLU activation function and saturation, which can both be represented exactly using hybrid zonotopes. The state-input map is shown in Figure \ref{fig:SUS_SinglePendulumOurs}(b). Using  \textbf{Theorem \ref{thm-FunctionCompSets}} and \textbf{Corollary \ref{co-affineDecomp}}, this took 4 seconds to compute and an additional 45 seconds to apply the exact complexity reduction techniques from~\cite{Bird_HybZono,BIRDthesis_2022}.

\subsubsection{Construction of closed-loop state-update set} \label{sec-ex1-phi}
 Given an over-approximation of the open-loop state-update set $\Bar{\Psi}$ and the exact state-input map $\Theta$, an over-approximation of the closed-loop state-update set $\Bar{\Phi}$ was constructed using the identity in \textbf{Theorem \ref{thm-CLSUSfromOL}} in less than 1 millisecond and exact reduction methods were completed in an additional 35 seconds. A projection of $\Bar{\Phi}$ is shown in Figure \ref{fig:SUS_SinglePendulumOurs}(c). This projection is a subset of the projection of $\Bar{\Psi}$ in Figure \ref{fig:SUS_SinglePendulumOurs}(a), as variability in the input is eliminated when creating the closed-loop state-update set using the state-input map $\Theta$. Although difficult to perceive in the figure, some thickness in the $x_{k,2}$ dimension remains as a result of the Taylor remainder $\mathcal{L}$ and over-approximating nonlinear functions of the plant model. Table~\ref{tab:ComplexitySets_Ex1} reports the memory complexity of $\bar{\Psi}$, $\Theta$, and $\bar{\Phi}$ for this example.

{
\subsubsection{Forward reachability} \label{sec-ex1-reach}
 Using \textbf{Corollary \ref{co-SusOA_ReachOA}} and iteration over the identity in \eqref{eq:1stepForward_CL}, over-approximations of forward reachable sets $\R_i$,  $i\in\{1,...,15\}$ are calculated from an initial set given by
\begin{align}
    \R_0 = \begin{bmatrix}
        -\pi, & \pi
    \end{bmatrix}\times\begin{bmatrix}
        -0.1, & 0.1
    \end{bmatrix}\:.
\end{align}%
The over-approximated reachable sets up to $\R_3$ are plotted in Figure \ref{fig:ReachSinglePendulumOurs}(a), overlaid by exact closed-loop trajectories found by randomly sampling points in $\X_0$ and propagating using \eqref{eqn-PendulumContinuous}. Examination of the exact trajectories exemplifies successful nonconvex over-approximation of the reachable sets. Computation time to execute the successor set identity was 5 milliseconds per time step on average. 

To handle growth in set memory complexity over time steps, set propagation methods often utilize over-approximations to reduce complexity. Using techniques from \cite{BIRDthesis_2022}, over-approximations of the reachable set are taken periodically every three time steps beginning at $k=3$, resulting in 4 total approximations that took an average of 49 seconds each to compute, in a manner similar to \cite{Siefert2021}. At the time steps corresponding to over-approximations, the set is first saved and analyzed before being over-approximated. The over-approximation is used to calculate the reachable set of the subsequent time step. Figure \ref{fig:ReachSinglePendulumOurs}(b) plots the over-approximated reachable sets and Figure \ref{fig:ReachSinglePendulumOurs}(c) plots the corresponding memory complexity. The periodic memory complexity reduction is apparent in Figure \ref{fig:ReachSinglePendulumOurs}(c). It is clear that the containment condition of \textbf{Theorem \ref{th:OneStep_F_CL}}, $\R_i\subseteq\domFX$, is met each time step. 

While plotting the reachable sets can be used to visually confirm performance and the containment condition of \textbf{Theorem \ref{th:OneStep_F_CL}}, this is a computationally expensive process, taking 7378 seconds to produce Figure \ref{fig:ReachSinglePendulumOurs}(b). Much of the same information can be obtained with lower computational burden by sampling the support function in the axis-aligned directions, which took a total of 34 seconds for all 15 steps, less than $0.5\%$ of the time to plot. 

\begin{figure}[htb!]
    \centering
    \begin{subfigure}[b]{2.75in}
         \centering
         \includegraphics[width=2.75in]{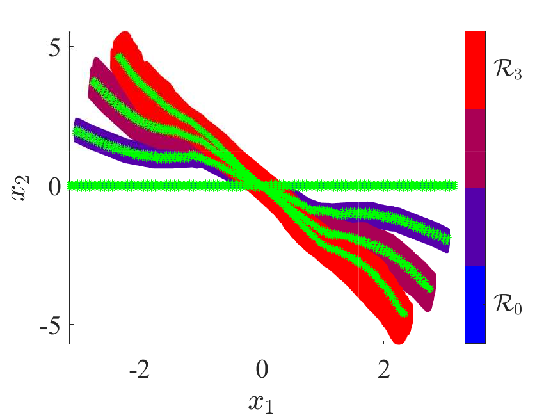}
         \caption{}
         \label{fig:ReachSinglePendulumOursNoTraj}
     \end{subfigure}\\
    \begin{subfigure}[b]{2.75in}
         \centering
         \includegraphics[width=2.75in]{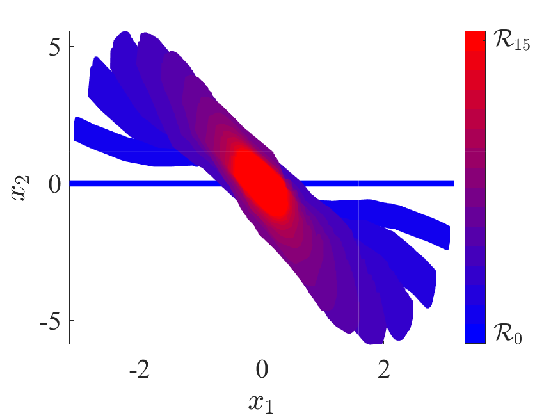}
         \caption{}
         \label{fig:ReachSinglePendulumOursWithTraj}
     \end{subfigure}\\
     \begin{subfigure}[b]{2.75in}
         \centering
         \includegraphics[width=2.75in]{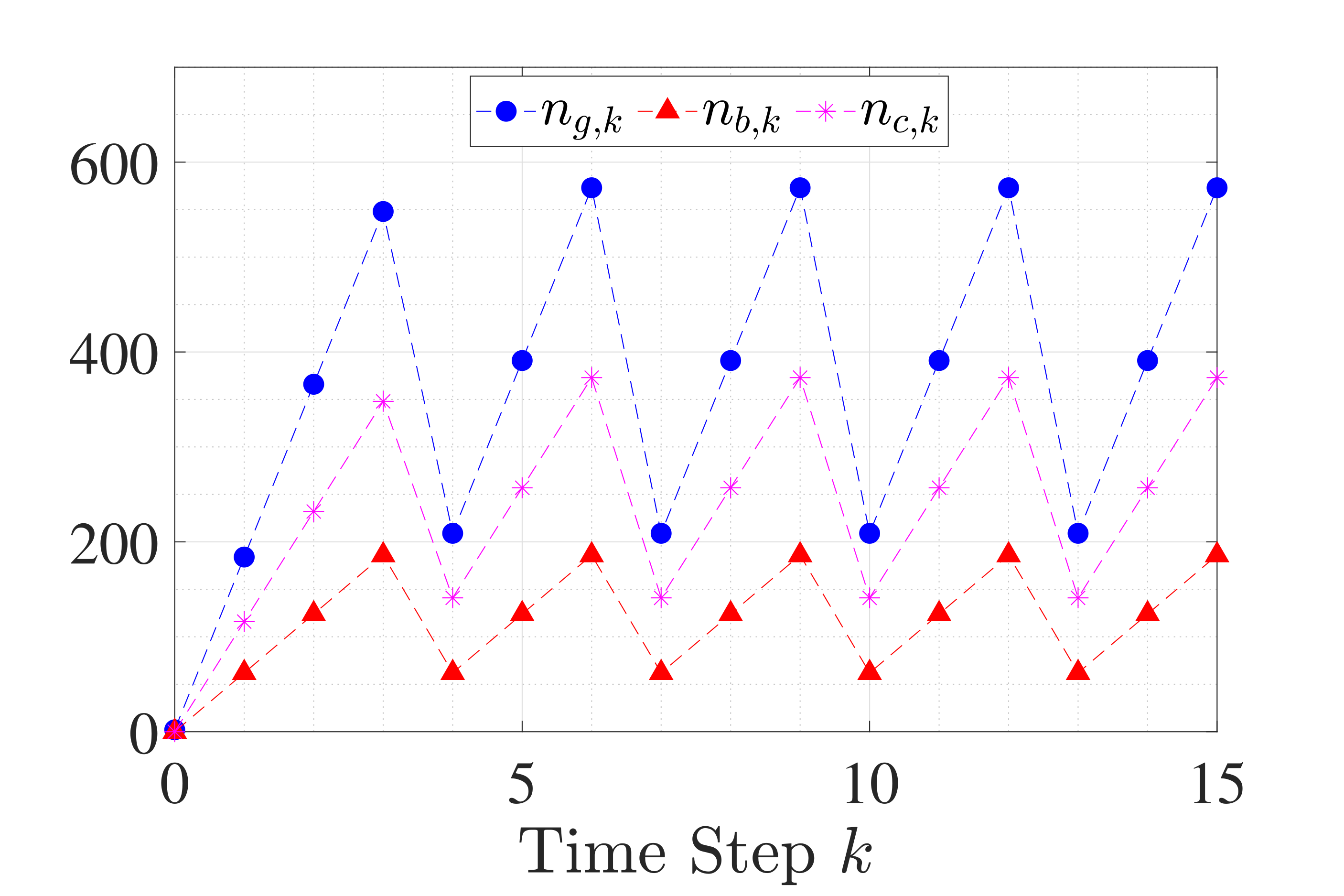}
         \caption{}
         \label{fig:complexity}
     \end{subfigure}
    \caption{(a) Over-approximation of reachable sets $\R_0\rightarrow\R_3$ of the inverted pendulum in closed-loop with a saturated neural network controller, overlaid by samples of exact trajectories in green. (b) Over-approximated reachable sets $\R_0\rightarrow\R_{15}$ with over-approximations taken every three time steps. (c) Memory complexity of the over-approximated reachable sets.}
    \label{fig:ReachSinglePendulumOurs}
    \vspace*{-\baselineskip}
\end{figure}
}

\subsection{ARCH AINNCS benchmark: Single pendulum}
\revM{R3.3,\\R5.2,\\R8.2} This section compares the proposed approach to the state-of-the-art using a modified version of the single pendulum benchmark in the Artificial Intelligence and Neural Network Control Systems (AINNCS) category of the Applied veRification for Continuous and Hybrid Systems (ARCH) workshop \cite[Section 3.5]{ARCH22:AINNCS} ,\cite[Section 3.5]{ARCH23:AINNCS}. This benchmark problem is similar to that presented in Section \ref{sec:SinglePendulmEx_OURS}, with different values for $g,l,m$ and a neural network with two hidden layers, each with 25 ReLU activation nodes. While the stated objective of the benchmark is to falsify or verify a safety condition, to demonstrate the proposed approach, this is modified to instead focus exclusively on calculating reachable sets over the time period $t\in\begin{bmatrix}
    0, & 1
\end{bmatrix}$ seconds with a time step of $0.05$ seconds. We consider a small initial set matching that in \cite[Section 3.5]{ARCH23:AINNCS} and a large initial set. These sets are given by
\begin{align}
\nonumber
    \text{Small initial set: } \R_0 &= \begin{bmatrix}
        1, & 1.2
    \end{bmatrix} \times \begin{bmatrix}
        0, & 0.2
    \end{bmatrix}\:, \text{ and}\\
\nonumber
    \text{Large initial set: } \R_0 &= \begin{bmatrix}
        0, & 1
    \end{bmatrix} \times \begin{bmatrix}
        -0.1, & 0.1
    \end{bmatrix}\:.
\end{align}

The proposed method follows the same procedure to construct the open-loop state-update set, state-input map, closed-loop state-update set, and reachable sets as done in the previous example. The complexity of the closed-loop state-update set is $n_{g,\phi}=321$, $n_{b,\phi}=75$, $n_{c,\phi}=240$. As done for the previous example, reachable sets are over-approximated every 3 time steps to handle growth in memory complexity.

The proposed method is compared to all four state-of-the-art tools that participated in the AINNCS category in 2022 and 2023, namely  CORA \cite{CORA}, JuliaReach \cite{JuliaReach}, NNV \cite{NNV}, and POLAR \cite{POLAR}. Computation times and an indication when reachable sets diverge with untenable over-approximation error for the small initial set and large initial set are given in Table~\ref{tab:PENDULUM_COMPTIME}. For the small initial set, all methods successfully generate reachable sets without diverging approximation error and the proposed method is substantially slower than the other tools. For the large initial set, the proposed method was the only method to compute the reachable set without diverging approximation error, though it required 40 minutes to do so.

\renewcommand{\arraystretch}{1.25} 
\begin{table}[]
\centering
\caption{Comparison of computation times in seconds of state-of-the-art tools for reachability analysis of an inverted pendulum with a neural network controller. Diverging approximation error is noted when experienced.}
\begin{tabular}{|c|c|c|}
\hline
Tool            & Small Initial Set & Large Initial Set                                                          \\ \hline
Proposed Method & 312               & 2377                                                                       \\ \hline
CORA            & 0.5               & \begin{tabular}[c]{@{}c@{}}0.6\\ Diverged\end{tabular}                     \\ \hline
JuliaReach      & 0.5               & \begin{tabular}[c]{@{}c@{}}0.7\\ Diverged\end{tabular}                     \\ \hline
NNV             & 2086              & \begin{tabular}[c]{@{}c@{}}\textgreater{}7200\\ Diverged\end{tabular}      \\ \hline
POLAR           & 0.2               & \begin{tabular}[c]{@{}c@{}}0.15\\ Terminated after \\ 7 steps\end{tabular} \\ \hline
\end{tabular}
\label{tab:PENDULUM_COMPTIME}
\end{table}
\renewcommand{\arraystretch}{1} 

CORA, JuliaReach, NNV, and POLAR each have tuning parameters that allow the user to adjust a trade-off between computation time and error. The results reported for the large initial set were generated with the same tuning parameters as used for the small initial set. The authors were unable to find alternative tuning parameters that avoided diverging over-approximating error for the large initial set. It is possible for CORA, JuliaReach, NNV, and POLAR to accommodate the large initial set via partitioning. For example, when the initial set is partitioned into 160 subsets, NNV can compute the reachable sets without diverging approximation error. In general, this may require a significant number of partitions, especially for systems with many states, that must be tuned to the problem at hand. 

More than $96\%$ of the total computation time of the proposed method is spent calculating periodic convex over-approximations. The time to reduce the order of the closed-loop state-update set is greater than $3\%$ of the total computation time. The time spent constructing the open-loop state-update set, state-input map, closed-loop state-update set, and successor sets for all time steps is only 8 seconds. Periodic over-approximations are taken to limit the complexity of the resulting sets, which also limits the complexity of analyzing them. This motivates work to efficiently generate hybrid zonotope over-approximations with acceptable approximation error, though approximations of hybrid zonotopes are challenging due to their implicit nature \cite{BIRDthesis_2022}.

{\subsection{High-dimensional Boolean function}
\revM{R6.4} We adopt the following example from \cite{alanwar2023polynomial}. Consider the Boolean function with $x_i,u_i\in \{0,1\}^{20}$, $i\in\{1,2,3\}$
\begin{align}
\nonumber
    x_{1,k+1}&=u_{1,k}\vee(x_{2,k} \odot x_{1,k})\:,\\
    \label{eqn-BoolDyn}
    x_{2,k+1}&=x_{2,k}\odot(x_{1,k} \wedge u_{2,k})\:,\\
    \nonumber
    x_{3,k+1}&=x_{3,k}\mathrlap{\sim}\wedge(u_{2,k}\odot u_{3,k})\:,
\end{align}%
where $\vee$, $\odot$, $\wedge$, and $\mathrlap{\sim}\wedge$ denote the standard Boolean functions OR, XNOR, AND, and NAND, respectively. Boolean functions can easily be represented as hybrid zonotopes, e.g., the set representing OR
\begin{align}
    \left\{ \begin{bmatrix}
        s_1\\
        s_2\\
        s_1 \vee s_2
    \end{bmatrix} \Bigg|\ (s_1,s_2)\in \{0,1\}^2\right\} \:,
\end{align}%
is equivalently given as the set of four points
\begin{align}
\label{eqn-ORpoints}
    \left\{ \begin{bmatrix}
        0\\
        0\\
        0
    \end{bmatrix},\begin{bmatrix}
        1\\
        0\\
        1
    \end{bmatrix},\begin{bmatrix}
        0\\
        1\\
        1
    \end{bmatrix},\begin{bmatrix}
        1\\
        1\\
        1
    \end{bmatrix} \right\}\:,
\end{align}%
which can be converted to a hybrid zonotope by concatenating the points \eqref{eqn-ORpoints} into a matrix $V$, using the incidence matrix $M=\id_4$, and applying the identity given in \textbf{Theorem \ref{thm-SOS2HYBZONO}}. Note that the resulting set maps two inputs consisting of the four combinations of Boolean values 0 and 1 to a single output of a Boolean value. Using a functional decomposition and \textbf{Theorem \ref{thm-FunctionCompSets}}, an exact open loop state-update set $\Psi$ for \eqref{eqn-BoolDyn} can be generated. Generating $\Phi$ in HCG-rep, including order reduction, took 0.82 seconds. 
 
An initial set for $(x_{1,k}$,$\ x_{2,k}$,$\ x_{3,k})$ consists of 8 possible values. Input sets for $(u_{1,k}$,$\ u_{2,k}$, $\ u_{3,k})$ also consist of 8 possible values. Reachable sets are calculated for 30 time steps. Computation times are plotted in Figure~\ref{fig:HCGvPolyLog_CompTime} comparing the methods presented here using HCG-rep to those developed for polynomial logical zonotopes. Beyond 5 steps, the computation time for polynomial zonotopes is on the order of hours due to exponential growth in the set complexity resulting from iterative AND (and NAND) operations, consistent with the results given in \cite[Table 3]{alanwar2023polynomial}. Computation time using hybrid zonotopes and the proposed methods scales polynomially with time. 

\begin{figure}
    \centering
    \includegraphics[width=3in]{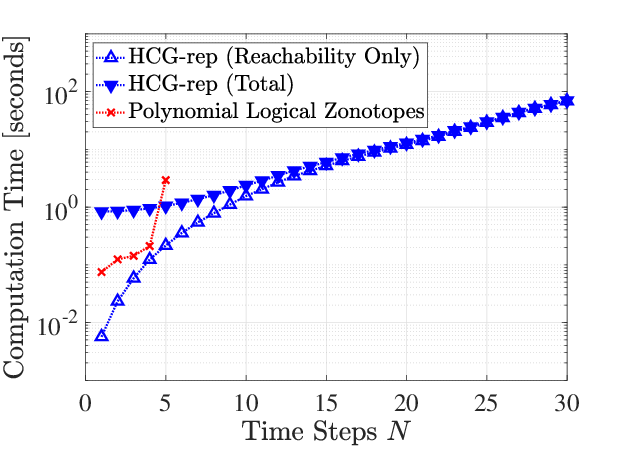}
    \caption{Computation time of reachable sets of a logical function. Two lines are plotted for HCG-rep, one including the time to generate the state-update set (total) and one that only includes the computation time associated with generation of reachable sets using \eqref{eq:1stepForward_CL}. The latter better demonstrates that computation time scales with $N$. Computation times for polynomial logical zonotopes are also shown.}
    \label{fig:HCGvPolyLog_CompTime}
\end{figure}

\subsection{ARCH AINNCS benchmark: Vertical collision avoidance system}

\revM{R3.3,\\R5.2,\\R8.2} As a second comparison to state-of-the-art reachability tools, we study the Vertical Collision Avoid System (VCAS) benchmark from the ARCH AINNCS category \cite{ARCH22:AINNCS,ARCH23:AINNCS}. The plant is a linear discrete-time model with $3$ states, given by
\begin{align}
    \nonumber
    h_{k+1} &= h_k - \dot{h}_{0,k} - \frac{1}{2} \ddot{h}_k\:,\\
    \nonumber
    \dot{h}_{0,k+1} &= \dot{h}_{0,k} + \ddot{h}_k\:,\\
    \nonumber
    \tau_{k+1} &= \tau_k - 1\:,
\end{align}
where $h_k$ is the relative height of the ownship from an intruder flying at a constant altitude, $\dot{h}_{0,k}$ is the derivative of the height of the ownship, and $\tau_k\in\{25,24,...,15\}$ is the time until the ownship and intruder are no longer horizontally separated. A controller state $\text{adv}_k\in\{1,2,...,9\}$, denotes the flight advisory, for which there are corresponding choices of $\ddot{h}$ to be selected by the pilot. At each time step, one of $9$ neural networks, each with 5 fully connected hidden layers and $20$ ReLU nodes per layer, is selected based on the previous time step advisory. Each neural network has $3$ inputs, corresponding to $h_k$, $\dot{h}_{0,k}$, and $\tau_k$, and has $9$ outputs, i.e., $f_{NN,i}(\cdot):\Rspace^3\rightarrow\Rspace^9\ \forall\ i\in\{1,...,9\}$. The index of the largest output of the neural network corresponding to the previous time step advisory determines the advisory for the current time step. If the previous advisory and current advisory coincide, and $\dot{h}_{0,k}$ complies with the advisory, then $\ddot{h}_k=0$. Otherwise, $\ddot{h}_{k}$ is selected according to the current advisory. The advisories, their associated ranges of compliant $\dot{h}_{0,k}$, and corresponding choices of $\ddot{h}_k$ are listed in Table~\ref{Tab:VCAS_Advisories}.

In \cite{ARCH22:AINNCS,ARCH23:AINNCS}, CORA, JuliaReach, and POLAR do not support reachability analysis of the closed-loop VCAS dynamics. Both CORA and JuliaReach provide custom simulation algorithms to falsify the VCAS benchmark, but do not employ reachability algorithms. NNV is the only tool to employ reachability algorithms for this benchmark problem and is able to verify/falsify various initial conditions by partitioning the set and omitting the multiple choices for $\ddot{h}$ given an advisory by assuming a ``middle'' (middle $\ddot{h}$ is chosen) or ``worst-case'' ($\ddot{h}$ that results in driving $h$ closest to $0$) strategy.

\renewcommand{\arraystretch}{1.4} 
\begin{table}[]
    \centering
    \caption{VCAS Advisories}
    \begin{tabular}{c|l|l|r}
       $\text{adv}$ & Advisory & Compliant $\dot{h}_{0,k}$ & Choice of $\ddot{h}_k$  \\ \hline \hline
       1 & COC & $\emptyset$ & $\{-\frac{g}{8},\ 0,\ \frac{g}{8}\}$ \\ \hline
       2 & DNC & $\dot{h}_{0,k} \leq 0$ & $\{-\frac{g}{3},\ -\frac{7g}{24},\ -\frac{g}{4}\}$ \\ \hline
       3 & DND & $0 \leq \dot{h}_{0,k}$ & $\{\frac{g}{4},\ \frac{7g}{24},\ \frac{g}{3}\}$ \\ \hline
       4 & DES1500 & $\dot{h}_{0,k} \leq -1500$ & $\{-\frac{g}{3},\ -\frac{7g}{24},\ -\frac{g}{4}\}$ \\ \hline
       5 & CL1500 & $1500 \leq \dot{h}_{0,k}$ & $\{\frac{g}{4},\ \frac{7g}{24},\ \frac{g}{3}\}$ \\ \hline
       6 & SDES1500 & $\dot{h}_{0,k} \leq -1500$ & $\{-\frac{g}{3}\}$ \\ \hline
       7 & SCL1500 & $1500 \leq \dot{h}_{0,k}$ & $\{\phantom{-}\frac{g}{3}\}$ \\ \hline
       8 & SDES2500 & $\dot{h}_{0,k} \leq -2500$ & $\{-\frac{g}{3}\}$ \\ \hline
       9 & SCL2500 & $2500 \leq \dot{h}_{0,k}$ & $\{\phantom{-}\frac{g}{3}\}$ \\ \hline
    \end{tabular}
    \label{Tab:VCAS_Advisories}
\end{table}
\renewcommand{\arraystretch}{1}

The proposed method is able to perform reachability analysis of this benchmark problem without this modification. We first generate a hybrid zonotope graph of the neural network associated with each of the 9 advisories. To reduce the complexity of the problem, it is first shown that only $4$ advisories, $\text{adv}_k\in\{1,5,7,9\}$, are achievable for a large region of the state space. Specifically, starting from an initial set where the previous advisory is $\text{COC},\ \text{adv}_{k-1} = 1$, only these advisories will occur for trajectories that remain within the constraints $h_k\in\begin{bmatrix}
    -400 & -100
\end{bmatrix}$ and $\dot{h}_{0,k}\in\begin{bmatrix}
    -100 & 100
\end{bmatrix}$. Compliant regions of advisories $4-9$ are never achieved within this domain. It will be shown that under these conditions, advisories $2$ and $3$ do not occur and can be neglected. 

The functional decomposition in Table~\ref{Tab:VCAS_FD_States2Adv} relates $h_k$, $\dot{h}_{0,k}$, $\tau_k$, and $\text{adv}_k$ to the next advisory, under the assumption that advisories 2 and 3 do not occur. To compactly write the decomposition, we use $\overrightarrow{w}_{i,j:k}$ to denote the vector of variables $\begin{bmatrix}
    w_{i,j} & w_{i,j+1} & \cdots & w_{i,k}
\end{bmatrix}^T$. Using the proposed methods and the functional decomposition, the graph of the function $\text{adv}_{k}=f(h_k,\dot{h}_{0,k},\tau_k,\text{adv}_{k-1})$ can be generated. Using the identity \eqref{eq:1stepForward_CL}, the set of advisories that can be active, given the set of states and previous advisory, is calculated. This is done iteratively in Table~\ref{tab:VCAS_ADVpropagation}. Iteration 4 results in the same potential advisories $\{1,5,7,9\}$ in the output set as the input set. The assumption that advisories $2$ and $3$ do not occur is confirmed, and the only advisories ever achieved starting from the input set in iteration 4, and remaining within the bounds for $(h_k,\dot{h}_k,\tau_k)$, is $\{1,5,7,9\}$. This knowledge allows the analysis that follows to neglect 5 of the neural networks from the closed-loop dynamics, significantly reducing the complexity of the problem over a large domain.

\renewcommand{\arraystretch}{1.1}
\begin{table}[htb]
\caption{Functional decomposition of VCAS: States \textrightarrow Advisory}
\centering
{\footnotesize
\begin{tabular}{lcl}
$w_1$                        & $=$ & $h_{k}$                                  \\
$w_2$                        & $=$ & $\dot{h}_{0,k}$                          \\
$w_3$                        & $=$ & $\tau_k$                                 \\
$w_4$                        & $=$ & $\text{adv}_{k-1}$                           \\
$\overrightarrow{w}_{i+4,1:9}$     & $=$ & 
$\begin{cases} 
f_{NN,i}(w_1,w_2,w_3), & \text{if }  \text{adv}_{k-1} = i\\
\mathbf{0}, & \text{otherwise} \end{cases}\:, \quad$\\
& & $\forall\ i\in\{1,2,3,4,5,6,7,8,9\}$\\
$\overrightarrow{w}_{14,1:9}$    & $=$ & $\sum_{i=5}^{13} \overrightarrow{w}_{i}$ \\
$\overrightarrow{w}_{15,1:7}$ & $=$ & $\begin{bmatrix}
    \max(w_{14,1},w_{14,2})\\
    \max(w_{15,1},w_{14,3})\\
    \max(w_{15,2},w_{14,4})\\
    \max(w_{15,3},w_{14,5})\\
    \max(w_{15,4},w_{14,6})\\
    \max(w_{15,5},w_{14,7})\\
    \max(w_{15,6},w_{14,8})
\end{bmatrix}$                              \\
$\overrightarrow{w}_{16,1:8}$ & $=$ & $\begin{bmatrix}
    w_{16,1} = \begin{cases}
        0, & \text{if } w_{14,1} \geq w_{14,2}\\
        1, & \text{if } w_{14,1} \leq w_{14,2}
    \end{cases}\phantom{a}\\
    w_{16,2} = \begin{cases}
        0, & \text{if } w_{15,1} \geq w_{14,3}\\
        1, & \text{if } w_{15,1} \leq w_{14,3}
    \end{cases}\phantom{a}\\
    \vdots\\
    w_{16,8} = \begin{cases}
        0, & \text{if } w_{15,7} \geq w_{14,9}\\
        1, & \text{if } w_{15,7} \leq w_{14,9}
    \end{cases}\phantom{a}
\end{bmatrix}$                              \\
$w_{17}$                     & $=$ & $\begin{cases}
    1, & \text{if } (w_{16,1:8}) = \mathbf{0}\\
    2, & \text{if } (w_{16,2:8}) = \mathbf{0}  \land w_{16,1} = 1\\
    3, & \text{if } (w_{16,3:8}) = \mathbf{0}  \land w_{16,2} = 1\\
    4, & \text{if } (w_{16,4:8}) = \mathbf{0}  \land w_{16,3} = 1\\
    5, & \text{if } (w_{16,5:8}) = \mathbf{0}  \land w_{16,4} = 1\\
    6, & \text{if } (w_{16,6:8}) = \mathbf{0}  \land w_{16,5} = 1\\
    7, & \text{if } (w_{16,7:8}) = \mathbf{0}  \land w_{16,6} = 1\\
    8, & \text{if } (w_{16,8:8}) = \mathbf{0}  \land w_{16,7} = 1\\
    9, & \text{if } \phantom{(w_{16,8:8}) = \mathbf{0}  \land \, \, \, } w_{16,8} = 1\\
\end{cases}$\\ 
\end{tabular}}
\label{Tab:VCAS_FD_States2Adv}
\end{table}
\renewcommand{\arraystretch}{1}

\begin{table}[]
    \centering
    \caption{Iterative domain propagation of advisories for an assumed domain of interest.}
    \label{tab:VCAS_ADVpropagation}
    {\footnotesize
    \begin{tabular}{c|c|c}
        Iteration & $\begin{array}{c}
             \text{Input Set}  \\
             (h_k,\dot{h}_k,\tau_k,\text{adv}_{k-1}) 
        \end{array}$ & $\begin{array}{c}
             \text{Reachable Advisories}  \\
             \text{adv}_{k} 
        \end{array}$ \\ \hline \hline
        1 &  $\begin{array}{l}
             [-400,\ -100]...\\
             \times[-100,\ 100]...\\
             \times\{25,24,...15\}...\\
             \times\{1\}
        \end{array}$ & $\{1,5\}$\\ \hline
        2 &  $\begin{array}{l}
             [-400,\ -100]...\\
             \times[-100,\ 100]...\\
             \times\{25,24,...15\}...\\
             \times\{1,5\}
        \end{array}$ & $\{1,5,7\}$\\ \hline
        3 &  $\begin{array}{l}
             [-400,\ -100]...\\
             \times[-100,\ 100]...\\
             \times\{25,24,...15\}...\\
             \times\{1,5,7\}
        \end{array}$ & $\{1,5,7,9\}$\\ \hline
        4 &  $\begin{array}{l}
             [-400,\ -100]...\\
             \times[-100,\ 100]...\\
             \times\{25,24,...15\}...\\
             \times\{1,5,7,9\}
        \end{array}$ & $\{1,5,7,9\}$\\
    \end{tabular}}
\end{table}

\renewcommand{\arraystretch}{1.1}
\begin{table}[]
\caption{Functional decomposition of VCAS: States \textrightarrow Updated States}
\centering
{\small
\begin{tabular}{lcl}
$w_1$                        & $=$ & $h_{k}$                                  \\
$w_2$                        & $=$ & $\dot{h}_{0,k}$                          \\
$w_3$                        & $=$ & $\tau_k$                                 \\
$w_4$                        & $=$ & $\text{adv}_{k-1}$                           \\
$\overrightarrow{w}_{i+4,1:4}$     & $=$ & 
$\begin{cases} 
[e_1^T,e_5^T,e_7^T,e_9^T]^T f_{NN,i}(w_1,w_2,w_3)\:,\\
\quad \text{if }  \text{adv}_{k-1} = i\\
\mathbf{0}\:, \quad \text{otherwise} \end{cases}$\\
& & $\forall\ i\in\{1,5,7,9\}$\\
$\overrightarrow{w}_{9,1:4}$    & $=$ & $\sum_{i=5}^{8} \overrightarrow{w}_{i}$ \\
$\overrightarrow{w}_{10,1:2}$ & $=$ & $\begin{bmatrix}
    \max(w_{9,1},w_{9,2})\\
    \max(w_{10,1},w_{9,3})\\
\end{bmatrix}$                              \\
$\overrightarrow{w}_{11,1:3}$ & $=$ & $\begin{bmatrix}
    w_{11,1} = \begin{cases}
        0, & \text{if } w_{9,1} \geq w_{9,2}\\
        1, & \text{if } w_{9,1} \leq w_{9,2}
    \end{cases}\phantom{a}\\
    w_{11,2} = \begin{cases}
        0, & \text{if } w_{10,1} \geq w_{9,3}\\
        1, & \text{if } w_{10,1} \leq w_{9,3}
    \end{cases}\phantom{a}\\
    w_{11,3} = \begin{cases}
        0, & \text{if } w_{10,2} \geq w_{9,4}\\
        1, & \text{if } w_{10,2} \leq w_{9,4}
    \end{cases}\phantom{a}
\end{bmatrix}$                              \\
$w_{12}$                     & $=$ & $\begin{cases}
    1, & \text{if } (w_{11,1:3}) = \mathbf{0}\\
    5, & \text{if } (w_{11,2:3}) = \mathbf{0}  \land w_{11,1} = 1\\
    7, & \text{if } (w_{11,3:3}) = \mathbf{0}  \land w_{11,2} = 1\\
    9, & \text{if } \hspace{2.22cm} w_{11,3} = 1\\
\end{cases}$\\
$w_{13}$ & $\in$ & $\begin{cases}
    \{-\frac{g}{8},0,\frac{g}{8}\}, & \text{if } w_{12} = 1\\
    \{\frac{g}{4},\frac{7g}{24},\frac{g}{3}\}, & \text{if } w_{12} = 5\\
    \{\frac{g}{3}\}, & \text{if } w_{12} = 7\\
    \{\frac{g}{3}\}, & \text{if } w_{12} = 9
\end{cases}$ \\
$w_{14}$ & $=$ & $w_1 - w_2 - w_{13}$ \\
$w_{15}$ & $=$ & $w_2 + w_{13}$ \\
$w_{16}$ & $=$ & $w_3 - 1$\\
\end{tabular}}
\label{Tab:VCAS_FD_SUS}
\end{table}
\renewcommand{\arraystretch}{1}

Now consider the domain $(h_k,\dot{h}_{0,k},\tau_k,\text{adv}_k)\in\text{D}_1\times\text{D}_2\times\text{D}_3\times\text{D}_4 = \begin{bmatrix}
    -400 & -100
\end{bmatrix}\times\begin{bmatrix}
    -100 & 100
\end{bmatrix}\times \{25,24,...,15\}\times\{1,5,7,9\}$. Using this domain and the functional decomposition given by Table~\ref{Tab:VCAS_FD_SUS}, a closed-loop state-update set is generated in $2.8$ seconds with complexity $(n_{g,\phi},n_{b,\phi},n_{c,\phi})=(2339,842,2049)$,
which encodes the transition from $(w_1,w_2,w_3,w_4)\rightarrow(w_{14},w_{15},w_{16},w_{12})$, i.e., from $(h_k,\dot{h}_{0,k},\tau_k,\text{adv}_k)$ to  $(h_{k+1},\dot{h}_{0,k+1},\tau_{k+1},\text{adv}_{k+1})$. Reachable sets are generated by recursion of \eqref{eq:1stepForward_CL} and checked for falsification each step, taking a total of 0.8 seconds to falsify the VCAS benchmark. Figure~\ref{fig:VCAS_reachFull_0to2} plots the resulting reachable sets.

\begin{figure}[htb!]
    \centering
    \includegraphics[width=3.4in]{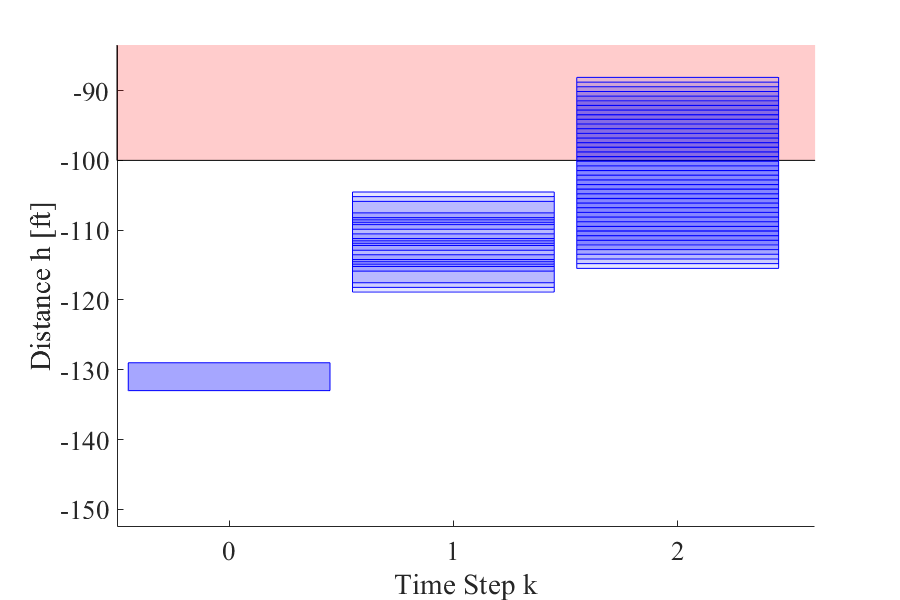}
    \caption{Reachable sets for the VCAS benchmark, calculated and falsified in $0.8$ seconds, and plotted in 4.3 seconds.}
    \label{fig:VCAS_reachFull_0to2}
\end{figure}

 \section{Conclusion}
\label{sect:conclusion}
{
This paper studies the reachability analysis of nonlinear systems, with focus on methods for efficiently constructing sets containing nonlinear functions. By leveraging the hybrid zonotope set representation and state-update sets, the proposed methods provide a unified framework for scalable calculation of reachable sets and their approximations spanning broad classes of systems including hybrid, logical, and nonlinear dynamics. The resulting reachable set memory complexity grows linearly with time and scales linearly in computational complexity with the state dimension. Numerical results demonstrate efficient computation and tight over-approximation of discrete-time reachable sets for a continuous-time nonlinear system in closed-loop with a neural network controller, scalability for a high-dimensional logical system, and exact and efficient reachability of a complicated vertical collision avoidance system that combines 9 neural networks with logic-based rules.
}

\bibliographystyle{IEEEtran}
\bibliography{bibNew}

\section*{Acknowledgements}{The authors thank Matthias Althoff, Tobias Ladner, Taylor Johnson, Diego Lopez, Christian Schilling, and Xin Chen for their insight regarding the implementation of CORA, NNV, JuliaReach, and POLAR for the ARCH inverted pendulum benchmark. This work was supported by the Department of Defense through the National Defense Science \& Engineering Graduate Fellowship Program.}

\vskip -2\baselineskip plus -1fil
\begin{IEEEbiography}[{\includegraphics[width=1in,height =1.25in,clip,keepaspectratio]{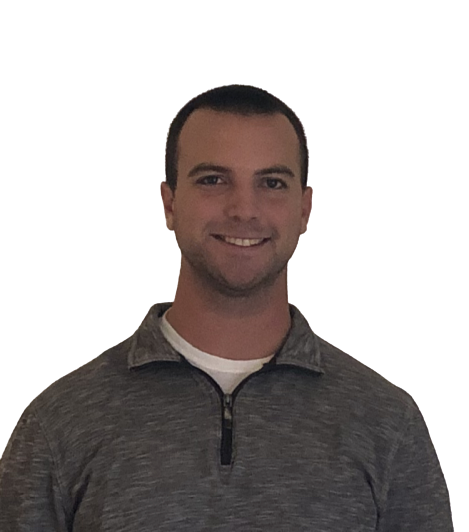}}]{Jacob Siefert} is a Ph.D. student and Research Assistant at The Pennsylvania State University, where he studies verification of advanced controllers using set-theoretic methods. He received his B.S. degree in Mechanical Engineering from the University of Maryland in 2016, and his M.S. degree in Mechanical Engineering from the University of Minnesota in 2021. His research interests include optimal control, co-design, hybrid systems, and reachability-based verification.
\end{IEEEbiography}
\vskip -2\baselineskip plus -1fil
\begin{IEEEbiography}[{\includegraphics[width=1in,height =1.25in,clip,keepaspectratio]{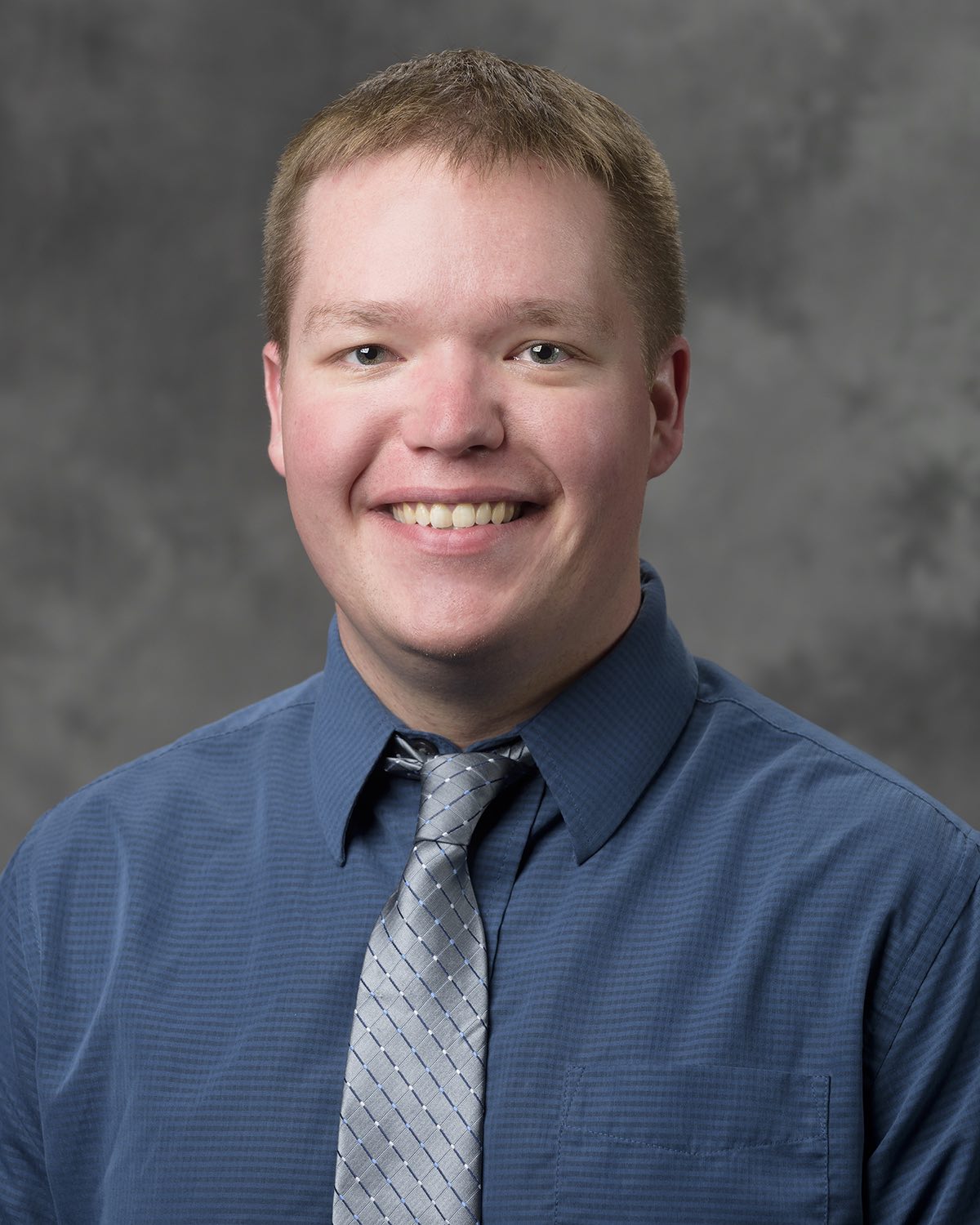}}]{Trevor J. Bird} is a Senior Lead Engineer at P. C. Krause and Associates, where he develops tools for the analysis, design, and optimization of transient systems. He received his B.S. degree in Mechanical Engineering from Utah State University in 2017, and went on to complete his M.S. and Ph.D. degrees in Mechanical Engineering from Purdue University in 2020 and 2022, respectively. His research interests include optimization, discrete mathematics, and set-based methods.  
\end{IEEEbiography}
\vskip -2\baselineskip plus -1fil
\begin{IEEEbiography}[{\includegraphics[width=1in,height =1.25in,clip,keepaspectratio]{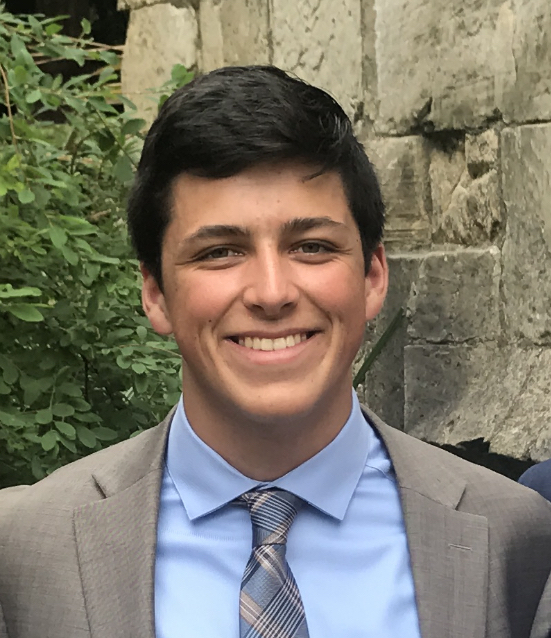}}]{Andrew F. Thompson} is a Ph.D. student and Research Assistant at The Pennsylvania State University. He received B.S. degrees in Mechanical Engineering and Computer Science from The University of Delaware in 2021, and an M.S. degree in Mechanical Engineering from The Pennsylvania State University in 2023. His research interests include trajectory optimization, reachability analysis, and applications to thermal systems, hybrid electric aircraft, and high-speed vehicles.
\end{IEEEbiography}
\vskip -2\baselineskip plus -1fil
\begin{IEEEbiography}[{\includegraphics[width=1in,height =1.25in,trim={0.5in 2.5in 0.5in 0.5in},clip,keepaspectratio]{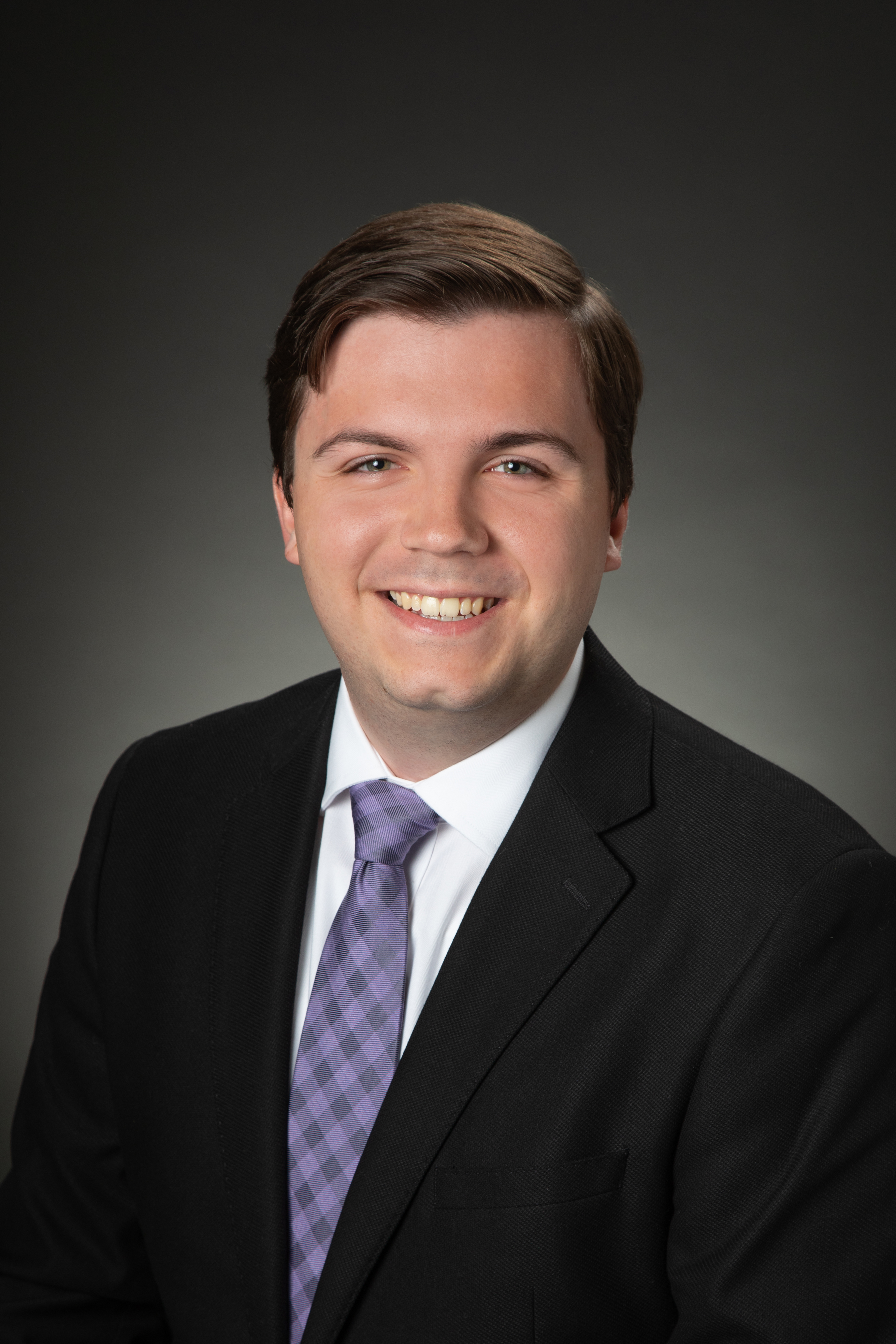}}]{Jonah G. Glunt} is a Ph.D. student and Research Assistant at The Pennsylvania State University, where he also received his B.S. degree in Mechanical Engineering in 2022. He is a recipient of the National Defense Science \& Engineering Graduate Fellowship. His research interests include optimization, autonomy, and set-theoretic methods for verified control and path planning.
\end{IEEEbiography}
\vskip -2\baselineskip plus -1fil
\begin{IEEEbiography}[{\includegraphics[width=1in,height =1.25in,clip,keepaspectratio]{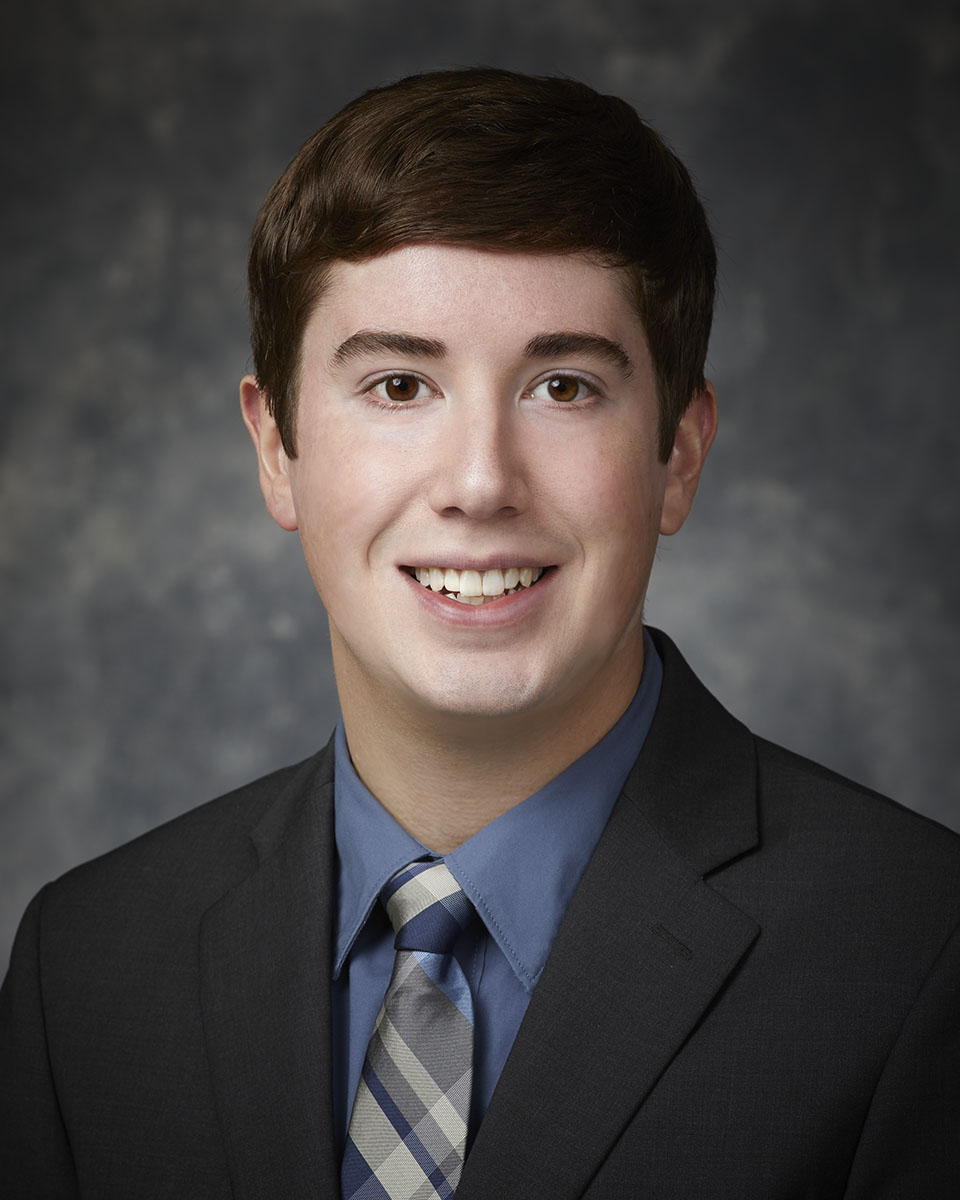}}]{Justin P. Koeln} received his B.S. degree in 2011 from Utah State University in Mechanical and Aerospace Engineering. He received M.S. and Ph.D. degrees in 2013 and 2016, respectively, from the University of Illinois at Urbana–Champaign in Mechanical Science and Engineering. He is an Assistant Professor at the University of Texas at Dallas in the Mechanical Engineering Department. He was a NSF Graduate Research Fellow and a Summer Faculty Fellow with the Air Force Research Laboratory. He was a recipient of the 2022 Office of Naval Research Young Investigator Award. His research interests include dynamic modeling and control of thermal management systems, model predictive control, set-based methods, and hierarchical and distributed control for electro-thermal systems.
\end{IEEEbiography}
\vskip -2\baselineskip plus -1fil
\begin{IEEEbiography}[{\includegraphics[width=1in,height =1.25in,clip,keepaspectratio]{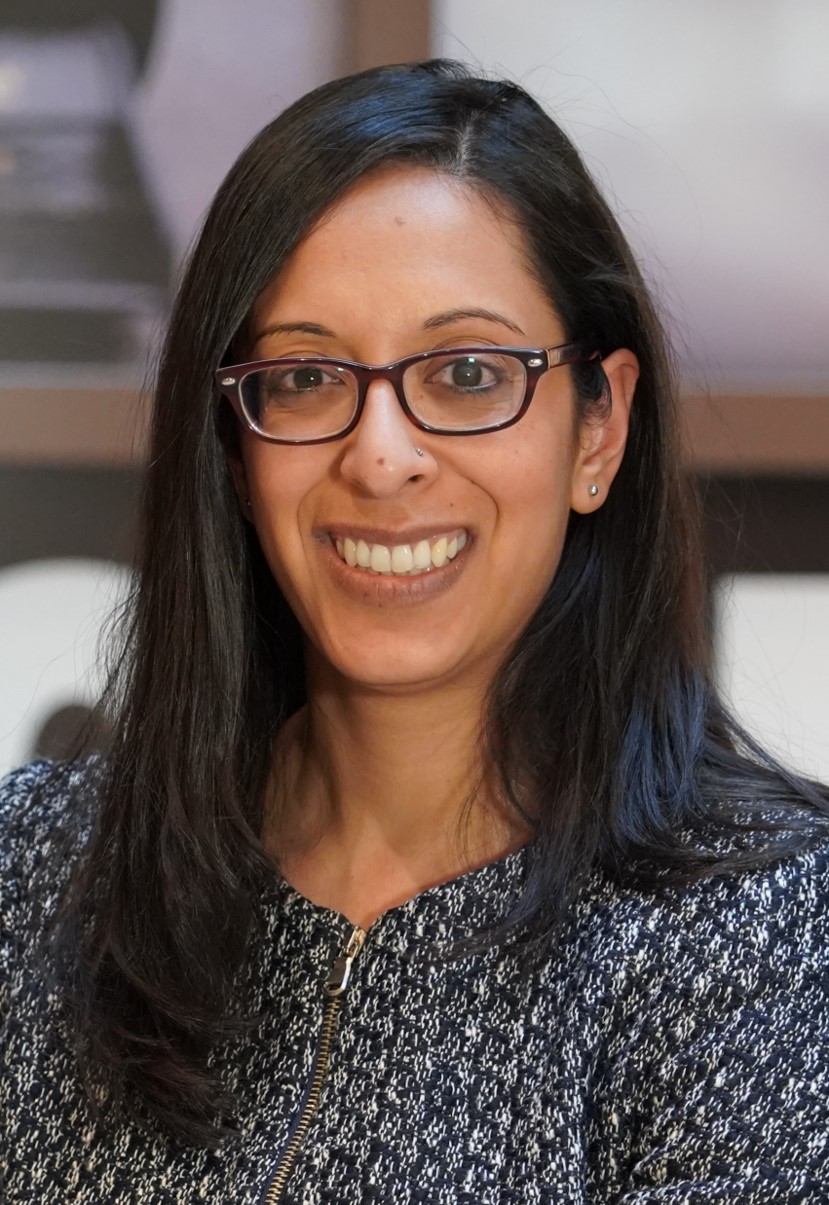}}]{Neera Jain} received the S.B. degree in Mechanical Engineering from the Massachusetts Institute of Technology in 2006. She received M.S. and Ph.D. degrees in 2009 and 2013, respectively, from the University of Illinois at Urbana–Champaign in Mechanical Science and Engineering. She is an Associate Professor of Mechanical Engineering at Purdue University. She is a recipient of the National Science Foundation CAREER Award (2022) and served as a National Research Council Senior Research Associate at the Air Force Research Laboratory (2022-2023). Her research interests include dynamic modeling, optimal control, and control co-design for complex energy systems and human-machine teaming.
\end{IEEEbiography}
\vskip -2\baselineskip plus -1fil
\begin{IEEEbiography}[{\includegraphics[width=1in,height =1.25in,clip,keepaspectratio]{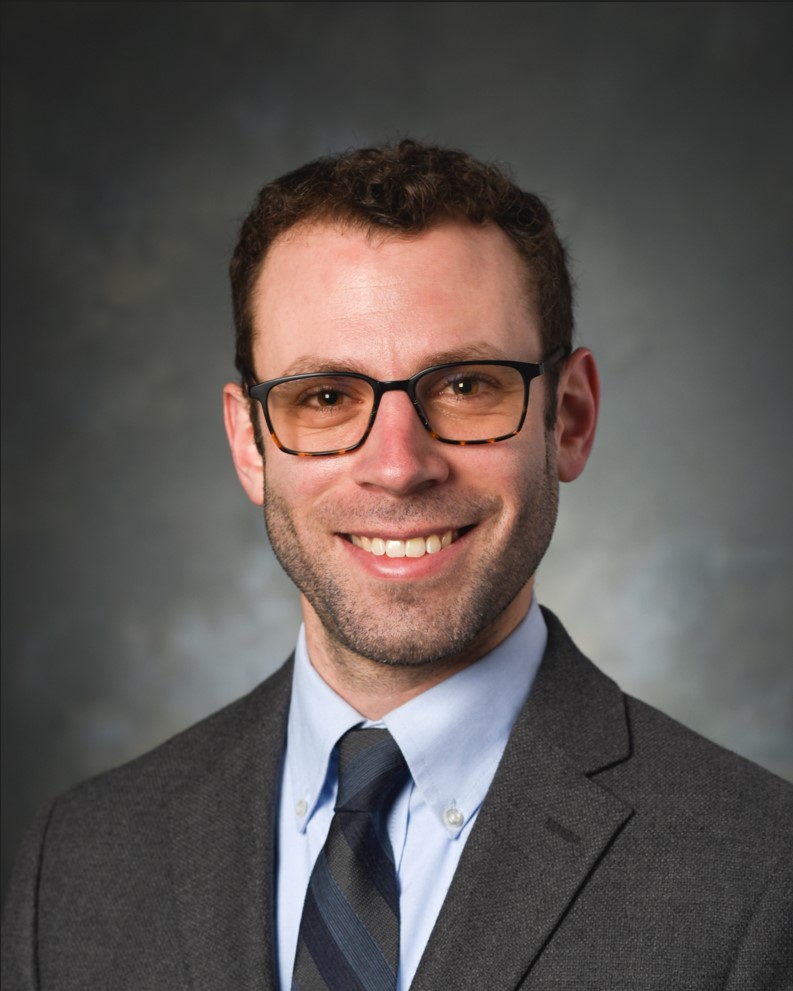}}]{Herschel C. Pangborn} received the B.S. degree in Mechanical Engineering from The Pennsylvania State University in 2013 and the M.S. and Ph.D. degrees in Mechanical Engineering from the University of Illinois at Urbana–Champaign in 2015 and 2019, respectively. He was an NSF Graduate Research Fellow and a Postdoctoral Research Associate at the University of Illinois, and a Summer Faculty Fellow with the Air Force Research Laboratory. He is currently an Assistant Professor with the Department of Mechanical Engineering and the Department of Aerospace Engineering (by courtesy) at The Pennsylvania State University. His research interests include model predictive control, optimization, and set-based verification of dynamic systems, including electro-thermal systems in vehicles and buildings.
\end{IEEEbiography}

\end{document}